\documentstyle[12pt,epsfig,amsfonts]{article}
\def \sech{\mathop{\rm sech}\nolimits}

\newcommand{\beq}{\begin{eqnarray}}
\newcommand{\eeq}{\end{eqnarray}}
\newcommand{\eqn}{\begin{equation}}
\newcommand{\een}{\end{equation}}
\begin{document}
\title{Generalized Zeta Functions and One-loop Corrections to Quantum Kink Masses}
\author{A. Alonso Izquierdo$^{(a)}$, W. Garc\'{\i}a Fuertes$^{(b)}$,
M.A. Gonz\'alez Le\'on$^{(a)}$ \\ and J. Mateos Guilarte$^{(c)}$
\\ {\normalsize {\it $^{(a)}$ Departamento de Matem\'atica
Aplicada}, {\it Universidad de Salamanca, SPAIN}}\\{\normalsize
{\it $^{(b)}$ Departamento de F\'{\i}sica}, {\it Universidad de
Oviedo, SPAIN}}\\ {\normalsize {\it $^{(c)}$ Departamento de
F\'{\i}sica}, {\it Universidad de Salamanca, SPAIN}}}

\date{}
\maketitle
\begin{abstract}
A method for describing the quantum kink states in the
semi-classical limit of several (1+1)-dimensional field
theoretical models is developed. We use the generalized zeta
function regularization method to compute the one-loop quantum
correction to the masses of the kink in the sine-Gordon and cubic
sinh-Gordon models and another two ${\rm P}(\phi)_2$ systems with
polynomial self-interactions.
\end{abstract}
\clearpage
\section{Introduction}

BPS states arising both in extended supersymmetric gauge theories
\cite{Olive} and string/M theory \cite{Duff} play a crucial r\^ole
in the understanding of the dualities between the different
regimes of the system. In this framework, domain walls appear as
extended states in ${\cal N}=1$ SUSY gluodynamics and the
Wess-Zumino model \cite{Dvali}. This circumstance prompted the
question of whether or not these topological defects saturate the
quantum Bogomolny bound. A return to the study of quantum
corrections to the masses of (1+1)-dimensional solitons has thus
been unavoidable. These subtle matters were first addressed in the
classical papers of Dashen, Hasslacher and Neveu, \cite{Dashen}, and Faddeev and Korepin, \cite{Kor},
for the purely bosonic $\lambda [\phi]^4_2$ and sine-Gordon
theories, and then in Reference \cite{Adda} for the
super-symmetric extension of these theories. Analysis of the
ultraviolet cut-off regularization procedure in the presence of a
background is the main concern in the papers of Reference
\cite{Rebhan}: the authors carefully distinguish between using a
cut-off either in the energy or in the number of modes. The second
method leads to the same result as in the computation performed by
DHN for bosonic fluctuations. Another point of view is taken in
Reference \cite{Shifman}, where SUSY boundary conditions related
more to infrared behaviour, are carefully chosen. On this basis,
and by using higher-derivative ultraviolet regularization (SUSY
preserving), the authors demonstrated an anomaly in the central
charge that compensates for the extra (quantum) contribution to
the classical mass.

In this paper we shall confine ourselves to purely bosonic
theories and leave the treatment of fermionic fluctuations for
future research. We address the quantization of non-linear waves
relying on the generalized zeta function regularization method to
control the infinite quantities arising in the quantum theory.
This procedure has been used previously in computing Casimir
energies and the quantum corrections to kink masses, see
\cite{Bordag}. We shall develop this topic, however, in a
completely general way, also offering a comparison with other
approaches. As well as obtaining exact results, we also shall
explain how asymptotic methods lead to a very good approximation
of the right answer. We believe that the novel application of the
asymptotic method should be very useful in the cubic sinh-Gordon
model as well as in multi-component scalar field theory, where the
traditional approach is limited by the lack of detailed knowledge
of the spectrum of the second-order fluctuation operator (see
\cite{AMG}, \cite{JMG} for extensive work on multi-component kinks
and their stability) .

The organization of the paper is as follows: In Section \S.2 the
general semi-classical picture of quantum solitons, the zeta
function regularization procedure, the zero-point energy and mass
renormalization prescriptions, and the asymptotic method are
described. In  Section \S.3, we apply the method to the \lq\lq
loop" kinks of the sine-Gordon , $\lambda (\phi)^4_2$, and cubic
sinh-Gordon models. In the first two paradigmatic cases, it is
possible to obtain an exact result, which allows comparison with
other methods. Approximate computations by means of the asymptotic
expansion of the heat function are also offered to test the
goodness of our procedure against the well known exact answers.
Section \S.4 is devoted to the analysis of the
 \lq\lq link" kink arising in the $\lambda (\phi)^6_2$ model.
Finally, Section \S.5 offers an outlook on further applications of
our approach.

\section{Semi-classical picture of quantum soliton states}

We shall consider (1+1)-dimensional scalar field theories whose classical
dynamics is governed by the action
\[
S[\psi]=\int d^2 y \left\{ \frac{1}{2} \,\frac{ \partial \psi}{ \,
\partial y^\mu} \frac{\partial \psi}{\partial y_\mu}-U({\psi}) \right\}
\]
We choose the metric tensor in $T^2({\Bbb R}^{1,1})$ as $g={\rm
diag}\,(1,-1)$ and the Einstein convention will be used throughout
the paper. We shall not use a natural unit system because we wish
to keep track of $\hbar$ in our formulas; nevertheless, we choose
the speed of light to be $c=1$. Elementary dimensional analysis
tells us that $[\hbar]=ML$, $[U(\psi)]=ML^{-1}$ and
$[\psi]=M^{\frac{1}{2}}L^{\frac{1}{2}}$ are the dimensions of the
important quantities.

The classical configuration space ${\cal C}$ is formed by the
static configurations ${\psi}(y)$, for which the energy functional
\[
E({\psi})=\int dy \left\{ \frac{1}{2} \frac{d \psi}{d y} \frac{d
\psi}{d y}+U({\psi}) \right\}
\]
is finite: ${\cal C}=\{{\psi}(y) \,/\, E({\psi}) < +\infty\}$. In
the Schr\"odinger picture, quantum evolution is ruled by the
functional differential Schr\"odinger equation:
\[
i \hbar \frac{\partial}{\partial t} \Psi[{\psi}(y),t]= {\Bbb H}
\Psi[{\psi}(y),t]\,.
\]
The quantum Hamiltonian operator
\[
{\Bbb H}=\int dy \left\{-\frac{\hbar^2}{2} \frac{\delta}{\delta
\psi(y)} \frac{\delta}{\delta \psi(y)}+E[{\psi}(y)] \right\}
\]
acts on wave functionals $\Psi[{\psi}(y),t]$ that belong to
$L^2({\cal C})$.

We wish to compute the matrix element of the evolution operator in
the \lq\lq field" representation
\begin{equation}
G\left(
{\psi}^{(f)}(y),{\psi}^{(i)}(y);T \right)=\left< {\psi}^{(f)}(y)
\right| e^{-\frac{i}{\hbar} T{\Bbb H}} \left|{\psi}^{(i)}(y)
\right>=\int {\cal D}[{\psi}(y,t)] \exp \left\{ -\frac{i}{\hbar}
S[{\psi}] \right\} \label{eq:ghat}
\end{equation}
for the choice
\[
{\psi}^{(i)}(y,0)={\psi}_{\rm K}(y) \hspace{0.4cm}, \hspace{0.4cm}
{\psi}^{(f)}(y,T)={\psi}_{\rm K}(y)
\]
where ${\psi}_{\rm K}(y)$ is a kink static solution of the
classical field equations. We are, however, only interested in the
loop ($\hbar $) expansion of ${G}$ up to the first quantum
correction. Also performing a analytic continuation to \lq\lq
Euclidean" time, $t=-i \tau$, $T=-i \beta$, this is achieved by
the steepest-descent method applied to the Feynman integral in
(\ref{eq:ghat}):
\[
G_{\rm E}\left( {\psi}_{\rm K}(y), {\psi}_{\rm K}(y);\beta \right)
\cong \exp \left\{ \rule{0cm}{0.6cm} \right. -\frac{\beta
E[{\psi}_{\rm K}]}{\hbar}  \left. \rule{0cm}{0.6cm}\right\}\, {\rm
Det}^{-\frac{1}{2}} \left[ \rule{0cm}{0.5cm} \right.
-\frac{\partial^2}{\partial \tau^2} +{\rm P}{\cal K} \left.
\rule{0cm}{0.5cm} \right] \left(1+o(\hbar)\right)
\]
where  ${\cal K}$
is the differential operator
\[
{\cal K}=-\frac{\partial^2}{\partial y^2}\, +\left. \frac{d^2 U}{d
\psi^2} \right|_{{\psi}_{\rm K}},
\]
and ${\rm P}$ is the projector over the strictly positive part of
the spectrum of ${\cal K}$. Note that, on the mathematical side,
the steepest-descent method provides a well defined approximation
to the Feynman integral if the spectrum of the quadratic form
${\cal K}$ is positive definite and, on the physical side, the
zero eigenvalue that appears in ${\rm Spec}({\cal K})$ contributes
to the next order in the loop expansion: it is due to neutral
equilibrium on the orbit of the kink solution under the action of
the spatial translation group. Moreover, in order to avoid the
problems that arise in connection with the existence of a
continuous spectrum of ${\cal K}$, we place the system in a
interval of finite but very large length $L$, i.e. $x\in
[-\frac{L}{2},\frac{L}{2}]$, and, eventually -after assuming
periodic boundary conditions on the small fluctuations all
throughout the paper- we shall let $L$ go to infinity.

>From the spectral resolution of ${\cal K}$,
\[
{\cal K} \xi_n(y)=\omega_n^2 \xi_n(y) \hspace{0.2cm},\hspace{1cm}
\omega_n^2\in {\rm Spec} ({\cal K})={\rm Spec} ({\rm P}{\cal
K})+\{0\} ,
\]
we write the functional determinant in the form
\[
{\rm Det}\, \left[ -\frac{\partial^2}{\partial \tau^2}
+{\cal K} \right] =\prod_n {\rm det} \,
\left[-\frac{\partial^2}{\partial \tau^2}+\omega_n^2 \right] .
\]
All the determinants in the infinite product correspond to
harmonic oscillators of frequency $i\omega_n$ and thus, with an
appropriate normalization, we obtain for large $\beta$
\[
G_{\rm E}\left( {\psi}_{\rm K}(y),{\psi}_{\rm K}(y);\beta\right) \cong
e^{-\frac{\beta}{\hbar} E[{\psi}_{\rm K}]} \prod_n \left(
\frac{\omega_n}{\pi \hbar} \right)^{\frac{1}{2}} \,
e^{-\frac{\beta}{2} \sum_n \omega_n (1+o(\hbar))}
\]
where the eigenvalue in the kernel of ${\cal K}$ has been
excluded.

Inserting eigen-energy wave functionals
\[
{\Bbb H} \Psi_j [{\psi}_{\rm K}(y)]=\varepsilon_j  \Psi_j
[{\psi}_{\rm K}(y)]
\]
we have an alternative expression for $G_{\rm E}$ for $\beta \rightarrow \infty$:
\[
G_{\rm E} \left( {\psi}_{\rm K}(y),{\psi}_{\rm K}(y);\beta\right)
\cong \Psi_0^{*}[{\psi}_{\rm K}(y)] \, \Psi_0[{\psi}_{\rm K}(y)]
\, e^{-\beta \frac{\varepsilon_0}{\hbar}}
\]
and, therefore, we obtain
\[
\varepsilon_0=E[{\psi}_{\rm K}]+\frac{\hbar}{2}
\sum_{\omega_n^2>0} \omega_n+o(\hbar^2)
\]
\[
\left| \Psi_0 [{\psi}_{\rm K}(y)] \right|^2={\rm
Det}^{\frac{1}{4}} \left[ \frac{P {\cal K}}{\pi^2
\hbar^2}\right]+o(\hbar^2) \hspace{0.2cm} ,
\]
as the kink ground state energy and wave functional up to one-loop
order.

We define the generalized zeta function
\[
\zeta_{P {\cal K}}(s)={\rm Tr}(P{\cal K})^{-s}=\sum_{\omega_n^2>0}
\frac{1}{(\omega_n^2)^s}
\]
associated to the differential operator ${\rm P}{\cal K}$. Then,
\begin{equation}
\varepsilon_0^{\rm K}=E[{\psi}_{\rm K}]+\frac{\hbar}{2} \, {\rm
Tr}\, (P{\cal K})^{\frac{1}{2}}+o(\hbar^2)=E[{\phi}_{\rm
K}]+\frac{\hbar}{2} \, \zeta_{P {\cal
K}}(-\textstyle\frac{1}{2})+o(\hbar^2) \label{eq:dosdos}
\end{equation}
\begin{equation}
\left| \Psi_0[{\psi}_{\rm K}(y)] \right|^2
=\pi\hbar\exp\left\{-\zeta_{P{\cal K}}(0)\right\} \exp\left\{
-\frac{1}{4} \frac{d \zeta_{P {\cal K}}}{ds}(0) \right\}
\label{eq:dostres}
\end{equation}
show how to read the energy and wave functional of the quantum
kink ground state in terms of the generalized zeta function of
the projection of the second variation operator ${\cal K}$ in the
semi-classical limit.

\subsection{The generalized zeta function regularization method :
zero-point energy and mass renormalizations }

The eigen-functions of ${\cal K}$ form a basis for the quantum
fluctuations around the kink background. Therefore, the sum of the
associated zero-point energies encoded in $\zeta_{P {\cal K}}
(-\frac{1}{2})$ in formula (\ref{eq:dosdos}) is infinite and we
need to use some regularization procedure. We shall regularize $
\zeta_{P {\cal K}} (-\frac{1}{2})$ by defining the analogous
quantity $ \zeta_{P {\cal K}} (s)$ at some point in the $s$
complex plane where $\zeta_{P {\cal K}}(s)$ does not have a pole.
$\zeta_{P {\cal K}}(s)$ is a meromorphic function of $s$, such
that its residues and poles can be derived from heat kernel
methods, see \cite{Gilkey}. If $K_{\rm {\cal K}}(y,z;\beta)$ is
the kernel of the heat equation associated with ${\cal K}$,
\begin{equation}
\left( \frac{\partial}{\partial \beta} +{\cal K} \right) K_{\rm
{\cal K}}(y,z;\beta)=0 \hspace{0.4cm}, \hspace{0.4cm} K_{\rm {\cal
K}}(y,z;0)=\delta(y-z) \label{eq:eqcalor}
\end{equation}
the Mellin transformation tells us that,
\[
\zeta_{P {\cal K}}(s)=\frac{1}{\Gamma(s)} \int_0^\infty d\beta \,
\beta^{s-1} \, h_{P {\cal K}} (\beta)
\]
where,
\[
h_{P{\cal K}}[\beta]={\rm Tr}\, e^{-\beta\, P {\cal K}}={\rm Tr}\,
e^{-\beta {\cal K}}-1 =-1+ \int dy \,K_{\rm {\cal K}}(y,y;\beta)
\]
is the heat function $h_{P {\cal K}}[\beta]$, if ${\cal K}$ is
positive semi-definite and ${\rm dim \, Ker}({\cal K})=1$. Thus,
the \lq\lq regularized" kink energy is in the semi-classical
limit:
\begin{equation}
\varepsilon_0^{\rm K}(s)=E[{\phi}_{\rm K}]+\frac{\hbar}{2} \mu^{2s+1}
\zeta_{P{\cal K}}(s)+ o(\hbar^2) \label{eq:ener}
\end{equation}
where $\mu$ is a unit of ${\rm length}^{-1}$ dimension, introduced
to make the terms in (\ref{eq:ener}) homogeneous from a
dimensional point of view. The infiniteness of the bare quantum
energy is seen here in the pole that the zeta function develops
for $s=-\frac{1}{2}$.

To renormalize $\varepsilon_0^{\rm K}$ we must: A. Subtract the
regularized vacuum quantum energy. B. Add counter-terms that will
modify the bare masses of the fundamental quanta, also regularized
by means of the generalized zeta function. C. Take the appropriate
limits.

A. The quantum fluctuations around the vacuum are governed by the
Schr\"odinger operator:
\[
{\cal V}=-\frac{d^2}{dy^2}+ \left. \frac{d^2 U}{d \psi^2}
\right|_{\psi_{\rm V}}
\]
where $\psi_{\rm V}$ is a constant minimum of $U[\psi]$. The
kernel of the heat equation
\[
\left( \frac{\partial}{\partial \beta}+{\cal V} \right)K_{\cal
V}(y,z;\beta)=0 \hspace{0.4cm}, \hspace{0.4cm} K_{\cal
V}(y,z;0)=\delta(y-z)
\]
provides the heat function $h_{\cal V}(\beta)$,
\[
h_{\cal V}(\beta)={\rm Tr}\, e^{-\beta {\cal V}}=\int \! dy \,
K_{\cal V}(y,y;\beta).
\]
We exclude the constant mode and, through the Mellin
transformation, we obtain
\[
\zeta_{\cal V}(s)=\frac{1}{\Gamma(s)} \int_0^\infty d\beta \,
\beta^{s-1}\, {\rm Tr} \, e^{-\beta {\cal V}}.
\]
The regularized kink energy measured with respect to the
regularized vacuum energy is thus:
\begin{eqnarray*}
\varepsilon^{\rm K}(s)&=&E[{\phi}_{\rm
K}]+\Delta_1\varepsilon^{\rm K}(s)+o(\hbar^2)\\&=&E[{\phi}_{\rm
K}]+\frac{\hbar}{2} \mu^{2s+1}\left[ \zeta_{P {\cal
K}}(s)-\zeta_{\cal V}(s)
 \right]+o(\hbar^2).
\end{eqnarray*}

B. If we now go to the physical limit $\varepsilon^{\rm
K}=\lim_{s\rightarrow -\frac{1}{2}} \varepsilon^{\rm K}(s)$, we
still obtain an infinite result. The reason for this is that the
physical parameters of the theory have not been renormalized. It
is well known that in (1+1)-dimensional scalar field theory normal
ordering takes care of all the infinities in the system: the only
ultraviolet divergences that occur in perturbation theory come
from graphs that contain a closed loop consisting of a single
internal line, \cite{Coleman}. From Wick's theorem, adapted to
contractions of two fields at the same point in space-time, we see
that normal ordering adds the mass renormalization counter-term,
\[
H(\delta m^2)=-\frac{\hbar}{2} \int \! dy  \, \delta m^2
:\frac{d^2 U}{d\psi^2}:+o(\hbar^2)
\]
to the Hamiltonian up to one-loop order. To regularize
\[
\delta m^2= \int \frac{dk}{4\pi}
\frac{1}{\sqrt{k^2+U^{\prime\prime}(\psi_{\rm V})}}
\]
we first place the system in a 1D box of length $L$ so that
$\delta m^2=\frac{1}{2L}\zeta_{{\cal V}}(\frac{1}{2})$, if the
constant eigen-function of ${\cal V}$ is not included in
$\zeta_{{\cal V}}$. Then, we again use the zeta function
regularization method and define: $\delta
m^2(s)=-\frac{1}{L}\frac{\Gamma (s+1)}{\Gamma (s)} \zeta_{{\cal
V}}(s+1)$. Note that $\delta m^2=\lim_{s\to -\frac{1}{2}}\delta
m^2(s)$. The criterion behind this regularization prescription is
the vanishing tadpole condition, which is shown in Appendix B of
Reference \cite{Bor} to be equivalent to the heat kernel
subtraction scheme.

The one-loop correction to the kink energy due to $H(\delta
m^2(s))$ is thus
\begin{eqnarray}
\Delta_2 \varepsilon^{\rm K}(s)&=&\left<\psi_{\rm K} \right|
H(\delta m^2(s)) \left| \psi_{\rm K} \right>-\left<\psi_{\rm V}
\right| H(\delta m^2(s)) \left| \psi_{\rm V} \right> = \nonumber\\
&=&
\lim_{L\rightarrow\infty}\frac{\hbar}{2L}\mu^{2s+1}\frac{\Gamma
(s+1)}{\Gamma (s)} \int_{-\frac{L}{2}}^{\frac{L}{2}} \!\! dy \,
\zeta_{{\cal V}}(s+1) \left[ \left. \frac{d^2 U}{d\psi^2}
\right|_{\psi_{\rm K}}- \left. \frac{d^2 U}{d\psi^2}
\right|_{\psi_{\rm V}} \right]=\nonumber
\\&=&-\lim_{L\rightarrow\infty}\frac{\hbar}{2L}\mu^{2s+1}\frac{\Gamma
(s+1)}{\Gamma (s)} \zeta_{{\cal
V}}(s+1)\int_{-\frac{L}{2}}^{\frac{L}{2}} \!\! dy \, V(y)\label{eq:deltados}
\end{eqnarray}
because the expectation values of normal ordered operators in
coherent states are the corresponding c-number-valued functions.

 C. The renormalized kink energy is thus
\begin{equation}
\varepsilon_R^{\rm K}=E[{\psi}_{\rm K}]+\Delta M_K
+o(\hbar^2)=E[{\psi}_{\rm K}]+ \lim_{s \rightarrow
-\frac{1}{2}}\left[ \Delta_1 \varepsilon^{\rm K}(s)+
\Delta_2\varepsilon^{\rm K}(s) \right]+o(\hbar^2)\label{eq:w1}
\end{equation}
whereas the renormalized wave functional reads
\begin{eqnarray*}
\int dx \left| \Psi_0^R({\psi}_{\rm K})\right|^2 &=& {\rm
Det}^{\frac{1}{4}} \left[ \frac{P {\cal K}}{\pi^2 \hbar^2} \right]
\, {\rm Det}^{\frac{(-1)}{4}} \left[ \frac{{\cal V}}{\pi^2
\hbar^2} \right]\\&=&\pi\hbar\exp\left\{-\left(\zeta_{P{\cal
K}}(0)-\zeta_{\cal V}(0)\right)\right\} \exp \left\{ -\frac{1}{4}
\left[ \frac{d \zeta_{P{\cal K}}}{d s}(0)-\frac{d \zeta_{\cal
V}}{ds}(0) \right] \right\}\,.
\end{eqnarray*}

\subsection{Asymptotic approximation to semi-classical kink
masses}

In order to use the asymptotic expansion of the generalized zeta
function of the ${\cal K}$ operator to compute the semi-classical
expansion of the corresponding quantum kink mass, it is convenient
to use non-dimensional variables. We define non-dimensional
space-time coordinates $x^\mu=m_dy^\mu$ and field amplitudes
$\phi(x^\mu)=c_d\psi(y^\mu)$, where $m_d$ and $c_d$ are constants
with dimensions $[m_d]=L^{-1}$ and
$[c_d]=M^{-\frac{1}{2}}L^{-\frac{1}{2}}$ to be determined in each
specific model. Also, we write
$\bar{U}(\phi)=\frac{c_d^2}{m_d^2}U(\psi )$.

The action and the energy can now be written in terms of their
non-dimensional counterparts:
\begin{eqnarray*}
S[\psi]&=&\frac{1}{c_d^2}\int d^2x\left(\frac{1}{2}\frac{\partial
\phi}{\partial x_\mu}\frac{\partial \phi}{\partial
x^\mu}-\bar{U}(\phi)\right)=\frac{1}{c_d^2}\bar{S}[\phi]\\
E[\psi]&=&\frac{m_d}{c_d^2}\int dx\left(\frac{1}{2}\frac{d
\phi}{dx}\frac{d
\phi}{dx}+\bar{U}(\phi)\right)=\frac{m_d}{c_d^2}\bar{E}[\phi]\,.
\end{eqnarray*}
The important point is that the Hessians at the vacuum and kink
configurations now read
\[
{\cal V}=m_d^2 \left(-\frac{d^2}{dx^2}+\bar{v}^2\right)=m_d^2 \,
\bar{\cal V}\hspace{0.5cm},\hspace{0.5cm}{\cal K}=m_d^2
\left(-\frac{d^2}{dx^2}+\bar{v}^2-\bar{V}(x)\right)=m_d^2\,\bar{\cal
K}
\]
where $\frac{d^2\bar{U}}{d\phi^2}|_{\phi_V}=\bar{v}^2$ and
$\frac{d^2\bar{U}}{d\phi^2}|_{\phi_K}=\bar{v}^2-\bar{V}(x)$.
Therefore,
\[
\zeta_{\cal V}(s)=\frac{1}{m_d^{2s}}\zeta_{\bar{\cal
V}}(s)\hspace{0.5cm},\hspace{0.5cm}\zeta_{\cal
K}(s)=\frac{1}{m_d^{2s}}\zeta_{\bar{\cal K}}(s)\,.
\]

The asymptotic expansion is superfluous if ${\rm Tr}\,e^{-\beta
P\bar{\cal K}}$ and $\zeta_{P\bar{\cal K}}(s)$ are susceptible of
an exact computation. If $\bar{V}(x)$ is a potential well of the
Posch-Teller type, see \cite{Drazin}, one can completely solve the
spectral problem for $\bar{\cal K}$ and there is no need for any
approximation to $\zeta_{P\bar{\cal K}}(s)$. In general the
spectrum of $\bar{\cal K}$ is not known in full detail, specially
in systems with multi-component kinks, and we can only determine
$\zeta_{P\bar{\cal K}}(s)$ by means of an asymptotic expansion.
Nevertheless, we shall also compute the asymptotic expansion of
$\zeta_{P\bar{\cal K}}(s)$ in the cases where the exact answer is
known in order to estimate the error accepted in this approach.

 In the formulas (\ref{eq:eqcalor}), (\ref{eq:ener}), (\ref{eq:deltados}) and (\ref{eq:w1}) we replace ${\cal V}$,
${\cal K}$ and $v^2$ by $\bar{\cal V}$, $\bar{\cal K}$ and
$\bar{v}^2$ and write the kernel of the heat equation for
$\bar{\cal K}$ in the form:
\[
K_{\bar{{\cal K}}}(x,x';\beta)=K_{\bar{{\cal V}}}(x,x';\beta)
A(x,x';\beta)\,;
\]
$A(x,x';\beta)$ is thus the solution of the PDE
\begin{equation}
\left( \frac{\partial}{\partial \beta}+\frac{x-x'}{\beta}
\frac{\partial}{\partial x}- \frac{\partial^2}{\partial
x^2}-\bar{V}(x) \right)A(x,x';\beta)=0 \label{eq:pde}
\end{equation}
with \lq\lq initial" condition: $A(x,x';0)=1$.

For $\beta<1$, we solve (\ref{eq:pde}) by means of an asymptotic
(high-temperature) expansion: $A(x,x';\beta)=\sum_{n=0}^\infty
a_n(x,x') \beta^n$. Note that there are no half-integer powers of
$\beta$ in this expansion because our choice of boundary
conditions with no boundary effects.

In this regime the heat function is given by:
\[
{\rm Tr}\, e^{-\beta \bar{\cal K}} = \int_{-\frac{m_d
L}{2}}^{\frac{m_dL}{2}} \hspace{-0.3cm} dx \, K_{\bar{\cal
K}}(x,x;\beta)=
\frac{e^{-\beta\bar{v}^2}}{\sqrt{4\pi\beta}}\sum_{n=0}^\infty
\int_{-\frac{m_d L}{2}}^{\frac{m_dL}{2}} \hspace{-0.3cm} dx \,
a_n(x,x)\beta^n=\frac{e^{-\beta \bar{v}^2
}}{\sqrt{4\pi\beta}}\sum_{n=0}^\infty a_n(\bar{\cal K}) \beta^n\,.
\]
It is not difficult to find the coefficients $a_n(x,x)$ by an
iterative procedure starting from $a_0(x,x')=1$. This procedure is
explained in the Appendix, which also includes the explicit
expression of some of the lower-order coefficients.

The use of the power expansion of $h_{P\bar{{\cal K}}}[\beta]={\rm
Tr}e^{-\beta \,P \bar{{\cal K}}}$ in the formula for the quantum
kink mass is quite involved: \hspace{0.2cm}

1. First, we write the generalized zeta function of $\bar{\cal V}$
in the form:
\[
\zeta_{\bar{\cal V}}(s)=\frac{1}{\Gamma(s)} \frac{m_d L}{\sqrt{4
\pi}} \int_0^1 d\beta \beta^{s-\frac{3}{2}} e^{-\beta\bar{v}^2}
+B_{\bar{\cal V}}(s)\,
\]
with
\[
B_{\bar{\cal V}}(s)=\frac{m_dL}{\sqrt{4 \pi}}
\frac{\Gamma[s-{\textstyle\frac{1}{2}},\bar{v}^2]}{\bar{v}^{2s-1}\Gamma[s]}
\hspace{0.4cm}, \hspace{0.4cm} \zeta_{\bar{\cal
V}}(s)=\frac{m_dL}{\sqrt{4\pi}}\frac{\gamma[s-{\textstyle\frac{1}{2}},\bar{v}^2]}{\bar{v}^{2s-1}\Gamma(s)}
+B_{\bar{\cal V}}(s)\,
\]
and $\Gamma[s,\bar{v}^2]$
and
$\gamma[s-\frac{1}{2},\bar{v}^2]$ being respectively the upper and lower incomplete gamma functions,
see \cite{Abramowitz}. It follows that $\zeta_{\bar{\cal V}}(s)$ is a meromorphic function of $s$
with poles at the poles of
$\gamma[s-{\textstyle\frac{1}{2}},\bar{v}^2]$, which occur when
$s-\frac{1}{2}$ is a negative integer or zero. $B_{\bar{\cal
V}}(s)$, however, is a entire function of $s$.

2. Second, from the asymptotic expansion of $h_{\bar{\cal
K}}[\beta]$ we estimate the generalized zeta function of
$P\bar{\cal K}$:
\begin{eqnarray*}
\zeta_{P\bar{\cal K}}(s)&=&\frac{1}{\Gamma(s)} \left[-\int_0^1
d\beta \beta^{s-1}+\frac{1}{\sqrt{4 \pi}} \sum_{n<n_0}
a_n(\bar{\cal K}) \int_0^1 d\beta \beta^{s+n-\frac{3}{2}}
e^{-\beta\bar{v}^2}+b_{n_0,\bar{\cal K}}(s) \right] +B_{P\bar{\cal
K}}(s)=\\ &=&-\frac{1}{s \Gamma(s)}+\frac{1}{\Gamma(s) \sqrt{4
\pi}} \sum_{n<n_0} a_n(\bar{\cal K})
\frac{\gamma[s+n-\frac{1}{2},\bar{v}^2]}{\bar{v}^{2(s+n-\frac{1}{2})}}+\frac{1}{\Gamma(s)}b_{n_0,\bar{\cal
K}}(s)+B_{P\bar{\cal K}}(s)
\end{eqnarray*}
where
\[
b_{n_0,\bar{\cal K}}(s)=\frac{1}{\sqrt{4\pi}} \sum_{n\geq
n_0}^\infty a_{n}(\bar{\cal K})
\frac{\gamma[s+n-\frac{1}{2},\bar{v}^2]}{\bar{v}^{2(s+n-\frac{1}{2})}}
\]
is holomorphic for ${\rm Re}\, s > -n_0+\frac{1}{2}$, whereas
\[
B_{P\bar{\cal K}}(s)= \frac{1}{\Gamma(s)}\int_{\bar{v}^2}
^\infty d\beta \,{\rm Tr} \, e^{-\beta\, P \bar{\cal K}}
\beta^{s-1}
\]
is a entire function of $s$. The values of $s$ where
$s+n-\frac{1}{2}$ is a negative integer or zero are the poles of
$\zeta_{P\bar{\cal K}}(s)$ because  the poles of
$\gamma[s+n-\frac{1}{2},\bar{v}^2]$ lie at these points in the
$s$-complex plane.

Renormalization of the zero-point energy requires the subtraction
of $\zeta_{\bar{\cal V}}(s)$ from $\zeta_{P \bar{\cal K}}(s)$. We
find,
\[
\zeta_{P\bar{\cal K}}(s)-\zeta_{\bar{\cal V}}(s) \approx
\frac{1}{\Gamma(s)} \left[ -\frac{1}{s}+\sum_{n=1}^{n_0-1}
\frac{a_n(\bar{\cal K})}{\sqrt{4 \pi}}
\frac{\gamma[s+n-\frac{1}{2},\bar{v}^2]}{\bar{v}^{2(s+n-\frac{1}{2})}}\right]
\]
and the error in this approximation with respect to the exact
result to $\Delta_1\varepsilon^K$ is:
\[
{\rm error}_1=\frac{\hbar m_d}{2}
[-\textstyle\frac{1}{2\sqrt{\pi}} \,b_{n_0,\bar{\cal
K}}(-{\textstyle\frac{1}{2}})+B_{P\bar{\cal
K}}(-{\textstyle\frac{1}{2}})-B_{\bar{\cal
V}}(-{\textstyle\frac{1}{2}})] .
\]
Note that the subtraction of $\zeta_{\bar{\cal V}}(s)$ exactly
cancels the contribution of $a_0(\bar{\cal K})$ and hence, the
divergence arising at $s=-\frac{1}{2}$, $n=0$. The quadratic
ultraviolet divergences appear in this scheme as related to the
pole of $\zeta_{\bar{\cal V}}(s)$ at $s=-\frac{1}{2}$, $n=0$.

\vspace{0.2cm}

3. Third, $\Delta_1\varepsilon^{\rm K}$ now reads
\begin{eqnarray*}
\Delta_1\varepsilon^{\rm K}&\cong & \frac{\hbar m_d}{\Gamma(-\frac{1}{2})}+ \frac{\hbar}{2} \lim_{s\rightarrow
-\frac{1}{2}} \left( \frac{\mu^2}{m_d^2}\right)^s \mu \,
\frac{a_1(\bar{\cal K})}{\sqrt{4\pi}\Gamma(s)}\frac{
\gamma[s+{\textstyle\frac{1}{2}},\bar{v}^2]}{\bar{v}^{2s+1}}+\\
&&+\frac{\hbar m_d}{2} \sum_{n=2}^{n_0-1} \frac{a_n(\bar{\cal
K})}{\sqrt{4 \pi}\Gamma(-\frac{1}{2})}
\frac{\gamma[n-1,\bar{v}^2]}{\bar{v}^{2n-2}}\,.
\end{eqnarray*}
The logarithmic ultraviolet divergences, hidden at first sight in
the DHN approach, arise here in connection with the pole of
$\zeta_{P\bar{\cal K}}(s)-\zeta_{\bar{\cal V}}(s)$ at
$s=-\frac{1}{2}$, $n=1$.

The surplus in energy due to the mass renormalization counter-term
is,
\begin{eqnarray*}
\Delta_2 \varepsilon^{\rm
K}&=&-\lim_{L\rightarrow\infty}\frac{\hbar \, a_1(\bar{\cal K})}{2
L} \lim_{s\rightarrow -\frac{1}{2}} \left( \frac{\mu}{m_d}
\right)^{2 s+1}\frac{\Gamma (s+1)}{\Gamma (s)} \zeta_{\bar{\cal
V}}(s+1) +o (\hbar^2\gamma)
\\ & \cong & -\frac{\hbar m_d}{2\sqrt{4\pi}}\, a_1(\bar{\cal K})
\lim_{s\rightarrow -\frac{1}{2}} \left( \frac{\mu}{m_d} \right)^{2
s+1}
\frac{\gamma[s+\frac{1}{2},\bar{v}^2]}{\bar{v}^{2s+1}\Gamma(s)}+o(\hbar^2\gamma)
\end{eqnarray*}
and the deviation from the exact result is
\[
{\rm error}_2= \lim_{L\rightarrow\infty}\frac{\hbar}{4L}
{a_1(\bar{\cal K})} B_{\bar{\cal V}}(\textstyle\frac{1}{2})\,.
\]
Therefore,
\begin{eqnarray*}
&&\varepsilon_R^{\rm K}=E[\psi_{\rm K}]+\Delta M_{\rm
K}+o(\hbar^2) \cong E[\psi_{\rm K}]-\frac{\hbar m_d}{2\sqrt{\pi}}
\left[ 1+\frac{1}{4\sqrt{\pi}} \sum_{n=2}^{n_0-1} a_n(\bar{\cal
K}) \frac{\gamma[n-1,\bar{v}^2]}{\bar{v}^{2n-2}}\right]+\\ & &
+\frac{\hbar m_d}{2\sqrt{4 \pi}} a_1(\bar{\cal
K})\lim_{s\rightarrow -\frac{1}{2}}\left(
\frac{\mu}{m_d}\right)^{2s+1} \left[
\frac{\gamma[s+\frac{1}{2},\bar{v}^2]}{\bar{v}^{2s+1}\Gamma(s)}-
\frac{\gamma[s+\frac{1}{2},\bar{v}^2]}{\bar{v}^{2s+1}\Gamma(s)}\right]+o(\hbar^2)\,.
\end{eqnarray*}
Note that the contributions proportional to $a_1(\bar{\cal K})$ of
the poles at $s=-\frac{1}{2}$ in $\Delta_1\varepsilon^K(s)$ and
$\Delta_2\varepsilon^K(s)$ cancel.

\vspace{0.2cm}

We are left with the very compact formula:
\begin{equation}
\Delta M_{\rm K}  \cong \hbar m_d \left[\Delta_0 +D_{n_0}\right]
\left\{ \displaystyle\begin{array}{l} \Delta_0 =
-\displaystyle\frac{1}{2\sqrt{\pi}} \\
D_{n_0}=-\displaystyle\sum_{n=2}^{n_0-1}\displaystyle\frac{a_n(\bar{\cal
K})}{8\pi}\displaystyle\frac{\gamma[n-1,\bar{v}^2]}{\bar{v}^{2n-2}}
\end{array}
\right. \label{asy}
\end{equation}
In sum, there are only two contributions to semi-classical kink
masses obtained by means of the asymptotic method: 1) $\hbar m_d
\Delta_0$ is due to the subtraction of the translational mode; 2)
$\hbar m_d D_{n_0}$ comes from the partial sum of the asymptotic
series up to the $n_0-1$ order. We stress that the merit of the
asymptotic method lies in the fact that there is no need to solve
the spectral problem of ${\cal K}$: all the information is encoded
in the potential $V(x)$.

\section{Loop kinks}

The existence of kinks is guaranteed if the minima of $U(\psi)$
are a discrete set which is the union of orbits of the discrete
symmetry group of the system. We shall use the term \lq\lq loop"
kinks to refer to those classical solutions that interpolate
between vacua belonging to the same orbit of the symmetry group;
otherwise, the solitary waves will be referred to as \lq\lq link"
kinks, see \cite{Boya}. In this Section we shall discuss three
kinks of the \lq\lq loop" type.

\subsection{The quantum sine-Gordon soliton}

We first treat the sine-Gordon model by considering the potential
energy density: $ U[\psi(y^\mu)]=\frac{m^4}{\lambda}(1-\cos
\frac{\sqrt{\lambda}}{m} \psi )$. The dimensions of the parameters
$m$ and $\lambda$ are respectively: $[m]=L^{-1}$ and
$[\lambda]=M^{-1}L^{-3}$. Therefore, we choose $m_d=m$ and
$c_d=\frac{\sqrt{\lambda}}{m}$ and find:
$\bar{U}[\phi(x,t)]=(1-\cos\phi)$.

The \lq\lq internal" symmetry group of the system is the infinite
dihedral group ${\bf D}_\infty={\Bbb Z}_2\times{\Bbb Z}$ generated
by internal reflections, $\phi\rightarrow -\phi$, and $2\pi$
translations, $\phi\rightarrow\phi+2\pi$. The vacuum classical
configurations $\phi_{\rm V}(x,t)=2 \pi n$ form the orbit ${\cal
M}=\frac{{\bf D}_\infty}{{\Bbb Z}_2}$ and there is spontaneous
symmetry breakdown of the internal translational symmetry through
the choice of vacuum. The moduli space of vacua, however,
$\hat{\cal M}=\frac{{\cal M}}{{\bf D}_\infty}$, is a single point
and all the equivalents kinks of the model,
\[
\phi_{\rm K}(x,t)=\pm 4\, \arctan e^x+2\pi n
\hspace{0.4cm},\hspace{0.4cm} \psi_{\rm K}(y,y_0)=\pm
\frac{4m}{\sqrt{\lambda}} \arctan e^{m y}+\frac{2 \pi
nm}{\sqrt{\lambda}}\hspace{0.4cm},
\]
are loop kinks. It is easy to check that $E[\psi_{\rm K}]=\frac{8
m^3}{\lambda}$ and $E[\psi_{\rm V}]=0$.

The second order variation operator around the kink solutions is
\[
{\cal K}=-\frac{d^2}{d y^2}+m^2 ( 1-2 \sech^2 m
y)\hspace{0.5cm},\hspace{0.5cm}\bar{{\cal K}}=-\frac{d^2}{d
x^2}+1-2 \sech^2 x\,.
\]
Note that ${\cal K}=m^2 \bar {\cal K}$; henceforth, $\zeta_{P{\cal
K}}(s)=\frac{1}{m^{2s}} \zeta_{P \bar{{\cal K}}}(s)$. Simili modo,
in the vacuum sector we have:
\[
{\cal V}=-\frac{d^2}{d y^2}+m^2 \hspace{0.4cm},\hspace{0.4cm}
\bar{{\cal V}}=-\frac{d^2}{dx^2}+1 \hspace{0.4cm},\hspace{0.4cm}
{\cal V}=m^2 \bar{{\cal V}}\hspace{0.4cm},\hspace{0.4cm}
\zeta_{\cal V}(s)=\frac{1}{m^{2s}} \zeta_{\bar{{\cal V}}}(s)\,.
\]

\subsubsection{Exact computation of the mass and the wave
functional}

\noindent$\bullet\ $Generalized zeta function of $\bar{{\cal V}}$:

The spectrum of $\bar{{\cal V}}$ acting on functions belonging to
$L^2({\Bbb R})$ is ${\rm Spec}\, \bar{{\cal V}}=k^2+1$, with $k\in
{\Bbb R}$ a real number. There is a half-bound state
$f_{k^2=0}(x)={\rm constant}$ that we shall not consider because
it is paired with the other half-bound state in Spec$(\bar{\cal
K})$. The spectral density on the interval $I=[-\frac{m
L}{2},\frac{m L}{2}]$  with periodic boundary conditions is
$\rho_{\bar{{\cal V}}}(k)=\frac{m L}{2 \pi}$. The heat function
is,
\[
{\rm Tr}\, e^{-\beta \bar{{\cal V}}}=\frac{mL}{2\pi}
\int_{-\infty}^\infty \!\! dk \, e^{-\beta(k^2+1)}=\frac{m
L}{\sqrt{4 \pi \beta}} \, e^{-\beta}
\]
and the generalized zeta function reads:
\[
\zeta_{\bar{{\cal V}}}(s)=\frac{mL}{\Gamma(s) \sqrt{4 \pi}}
\int_0^\infty \!\! d\beta \, \beta^{s-\frac{3}{2}}
e^{-\beta}=\frac{mL}{\sqrt{4 \pi}}
\frac{\Gamma(s-\frac{1}{2})}{\Gamma(s)}\,.
\]
Therefore, $\zeta_{\bar{{\cal V}}}(s)$ (hence $\zeta_{\cal V}(s)$)
is a meromorphic function of $s$ with poles at $s=\frac{1}{2}$,
$-\frac{1}{2}$, $-\frac{3}{2}$, $-\frac{5}{2}$, $\cdots$. The
generalized zeta function of the Hessian at the vacuum is,
however, also infrared-divergent: it is linearly divergent when $L
\rightarrow \infty$ even at points $s\in {\Bbb C}$ where
$\zeta_{\cal V}(s)$ is regular.

\noindent$\bullet\ $Generalized zeta function of $\bar{{\cal K}}$:

In this case ${\rm Spec}\, \bar{{\cal K}}=\{0\} \cup \{k^2+1\}$,
$k\in {\Bbb R}$ and the spectral density on $I$ is
\[
\rho_{\bar{{\cal K}}}(k)=\frac{mL}{2\pi}+\frac{1}{2\pi} \frac{d
\delta(k)}{d k}
\]
with phase shifts
\[
\delta(k)=2 \arctan \frac{1}{k}
\]
because $\bar{{\cal K}}$ is the Schr\"odinger operator that
governs the scattering through the first of the \lq\lq
transparent" Posch-Teller potentials, \cite{Drazin}. Thus,
\begin{eqnarray*}
{\rm Tr}\, e^{-\beta \bar{{\cal K}}}&=&1+\frac{mL}{2\pi}
\int_{-\infty}^\infty dk e^{-\beta(k^2+1)}+\frac{1}{2\pi}
\int_{-\infty}^\infty dk \frac{d\delta (k)}{dk}
e^{-\beta(k^2+1)}=\\ &=& 1+\frac{mL}{\sqrt{4\pi \beta}}
e^{-\beta}-{\rm Erfc}\, \sqrt{\beta}
\end{eqnarray*}
where ${\rm Erfc}\, \sqrt{\beta}$ is the complementary error
function, \cite{Abramowitz}. Note that $\bar{{\cal K}}$ has a zero
mode, the eigen-function being the translational mode $\frac{d
\phi_K}{dx}=\sech^2 x$, which  must be subtracted because it
arises in connection with the breaking of the translational
symmetry, $x\rightarrow x+a$, by the kink solution and does not
contribute to the kink mass up to this order in the loop
expansion. There is also a half-bound state, $f_{k^2=0}(x)={\rm
tanh}x$, that exactly cancels the contribution of the constant
half-bound state in ${\rm Spec}{\cal V}$. Therefore, we obtain
\[
{\rm Tr} e^{-\beta\, P \bar{{\cal K}}}={\rm Tr} e^{-\beta
\bar{{\cal K}}}-1={\rm Tr} e^{-\beta \bar{{\cal V}}}-{\rm Erfc}\,
\sqrt{\beta}
\]
and
\begin{displaymath}
\zeta_{P \bar{{\cal K}}}(s)= \zeta_{\bar{{\cal
V}}}(s)-\frac{1}{\Gamma(s)} \int_0^\infty \!\!\! d\beta \, {\rm
Erfc}\,\sqrt{\beta} \beta^{s-1}=\zeta_{\bar{{\cal
V}}}(s)-\frac{1}{\sqrt{\pi}} \frac{\Gamma(s+\frac{1}{2})}{s
\Gamma(s)}\,.
\end{displaymath}
$\zeta_{P\bar{{\cal K}}}(s)$ (hence $\zeta_{P{\cal K}}(s)$) is
also a meromorphic function of $s$ that shares all the poles with
$\zeta_{\bar{{\cal V}}}(s)$, but the residues are different except at
$s=\frac{1}{2}$, a pole where the residues of
$\zeta_{P\bar{{\cal K}}}(s)$ and $\zeta_{\bar{{\cal V}}}(s)$
coincide. The infrared divergence, however, is identical in the
kink background and the vacuum.

We can now compute the limit of the regularized quantities that
enter in the one-loop correction formula to the kink mass:
\begin{eqnarray}
\Delta_1\varepsilon^{\rm K}&=&\frac{\hbar}{2} \left[
\lim_{s\rightarrow -\frac{1}{2}} \left( \frac{\mu^2}{m^2}
\right)^s \mu (\zeta_{P \bar{{\cal K}}}(s)-\zeta_{\bar{{\cal
V}}}(s))\right] =-\frac{\hbar m}{2\sqrt{\pi}} \left[ \lim_{\varepsilon
\rightarrow 0} \left( \frac{\mu}{m} \right)^{2\varepsilon}
\frac{\Gamma (\varepsilon )}{\Gamma
(\frac{1}{2}+\varepsilon  )}\right ]\nonumber\\&=& \frac{\hbar
m}{2\pi}\lim_{\varepsilon\rightarrow 0}\left[
-\frac{1}{\varepsilon}-2\log\frac{2\mu}{m}-\psi(1)+\psi(\frac{1}{2})+o(\varepsilon
)\right ]
\end{eqnarray}
and
\begin{eqnarray*}
\Delta_2 \varepsilon^{\rm K} &=& -\frac{2\hbar}{L}
\lim_{s\rightarrow -\frac{1}{2}} \left( \frac{\mu}{m}
\right)^{2s+1}\frac{\Gamma (s+1)}{\Gamma (s)} \zeta_{\bar{{\cal
V}}}(s+1) + o(\hbar^2\gamma)=-\frac{\hbar m}{\sqrt{\pi}}\lim_{s\to
-\frac{1}{2}}\left(\frac{\mu}{m}\right)^{2s+1}\frac{\Gamma
(s+\frac{1}{2})}{\Gamma (s)}\\&=&-\frac{\hbar
m}{\sqrt{\pi}}\lim_{\varepsilon\to
0}\left(\frac{\mu}{m}\right)^{2\varepsilon}\frac{\Gamma
(\varepsilon)}{\Gamma (-\frac{1}{2}+\varepsilon)} = \frac{\hbar m}{2 \pi}
\lim_{\varepsilon \rightarrow 0} \left [
\frac{1}{\varepsilon}+2\log\frac{2\mu}{m}+\psi(1)-\psi(-\frac{1}{2})+o(\varepsilon)\right]+o(\hbar^2\gamma)\,.
\end{eqnarray*}
where $\psi(z)=\frac{\Gamma^\prime(z)}{\Gamma(z)}$ is the digamma
function.

The important point to notice is that the renormalization of the
zero-point energy performed by the subtraction of $\zeta_{\cal
V}(-\frac{1}{2})$ still leaves a divergence coming from the
$s=-\frac{1}{2}$ poles because the residues are different. The
correction due to the mass renormalization counter-term also has a
pole. The sum of the contributions of the two poles leaves a
finite remainder and we end with the finite answer:
\begin{equation}
\Delta_1\varepsilon^{\rm K}+\Delta_2 \varepsilon^{\rm
K}=-\frac{\hbar m}{\pi}\hspace{0.5cm},\hspace{0.5cm}
\varepsilon_R^{\rm K}=E[\psi_{\rm K}]-\frac{\hbar
m}{\pi}+o(\hbar^2\gamma)=\frac{8 m}{\gamma}-\frac{\hbar
m}{\pi}+o(\hbar^2\gamma) \label{eq:enersol}.
\end{equation}
The one-loop quantum correction to the mass of the sine-Gordon
soliton obtained by means of the generalized zeta function
procedure exactly agrees with the accepted result, see
\cite{Dashen}, \cite{Kor}, and, henceforth, with the outcome of
the mode number regularization method, \cite{Rebhan}.

The square of the modulus of the ground state wave functional up
to one-loop order is given by (\ref{eq:dostres}). If
$W=\frac{P\bar{{\cal K}}}{C^2}$ and
$C=\frac{\pi\hbar\lambda}{m_d^2}$, obviously, $\zeta_W(s)=C^{2s}
\zeta_{P{\bar {\cal K}}}(s)$ ($\zeta_W(0)=\zeta_{P{\bar{{\cal
K}}}}(0)$) and we have
$
\frac{d \zeta_W}{ds}=C^{2s} \zeta_{P{\bar {\cal K}}}(s) \log C
+C^{2s} \frac{d \zeta_{P\bar{{\cal K}}}}{ds}(s).
$
Thus,
\begin{eqnarray}
\frac{d \zeta_{P\bar{\cal K}}}{ds}&=& \frac{d \zeta_{\bar{\cal
V}}}{ds}-\frac{1}{\sqrt{\pi}} \frac{\Gamma(s+\frac{1}{2})}{s
\Gamma(s)}
\left[\psi(s+{\textstyle\frac{1}{2}})-\frac{1}{s}-\psi(s)\right]\nonumber\\
\hspace{1cm} \frac{d \zeta_{\bar{\cal V}}}{ds}&=& \frac{mL}{\sqrt{4
\pi}} \frac{\Gamma(s-\frac{1}{2})}{\Gamma(s)}
\left[\psi(s-{\textstyle\frac{1}{2}})-\psi(s)\right]
\end{eqnarray}
and
\[
\zeta_{\bar{\cal V}}(0)=0 \hspace{0.4cm}, \hspace{0.4cm}
\zeta_{P\bar{\cal K}}(0)=-1 \hspace{0.4cm}, \hspace{0.4cm} \frac{d
\zeta_{\bar{\cal V}}}{ds}(0)=-mL \hspace{0.4cm}, \hspace{0.4cm}
\frac{d \zeta_{P\bar{\cal K}}}{ds}(0)=-m L
+\psi(1)-\psi({\textstyle\frac{1}{2}}).
\]
Therefore,
\begin{displaymath}
\left| \Psi_0(\phi_{\rm K}(x))\right|^2 =\sqrt{ \frac{C}{2}}
\exp \left\{ \frac{1}{4} m L \right\},
\ \ \ \ \ \ \left|
\Psi_0(\phi_{\rm V}(x))\right|^2 = \exp \left\{ \frac{1}{4} m L
\right\}.
\end{displaymath}
Renormalizing the wave functional with respect to the vacuum we obtain
\begin{equation}
\frac{\left| \Psi_0(\phi_{\rm K}(x))\right|^2}{ \left|
\Psi_0(\phi_{\rm V}(x))\right|^2} =\sqrt{ \frac{C}{2}}
\label{eq:div}
\end{equation}

\subsubsection{The asymptotic expansion and quantum corrections}

In the sine-Gordon model the exact formulas for ${\rm
Tr}\,e^{-\beta \, P \bar{\cal K}}$ and $\zeta_{P \bar{\cal K}}(s)$
are readily derived because the spectrum of the Schr\"odinger
operator $\bar{{\cal K}}$ is completely known. On the other hand,
the series expansion of the complementary error function tells us
that
\[
{\rm Tr} e^{-\beta \, P  \bar{\cal K}}= \left[ \frac{mL}{\sqrt{4
\pi \beta}}+\frac{1}{\sqrt{\pi}} \sum_{n=1}^\infty \frac{2^n\,
\beta^{n-\frac{1}{2}}}{1\cdot 3\cdot 5 \cdot ... \cdot (2n-1)}
\right] e^{-\beta}-1
\]
and the $a_n(\bar{\cal K})$ coefficients can be computed from this
exact expression:
\[
{\rm Tr} e^{-\beta \, P \bar{\cal K}}
=\frac{e^{-\beta}}{\sqrt{4\pi}}\sum_{n=0}^\infty a_n(\bar{\cal K})
\beta^{n-\frac{1}{2}}-1 \hspace{0.4cm}, \hspace{0.4cm}
a_0(\bar{\cal K})=m L \hspace{0.4cm}, \hspace{0.4cm}
a_{n}(\bar{\cal K})=\frac{2^{n+1}}{(2n-1)!!}\,.
\]
One can check by direct calculation that indeed,
\[
a_n(\bar{\cal
K})=\int_{-\frac{mL}{2}}^{\frac{mL}{2}}\hspace{-0.3cm} dx\,
a_{n}(x,x)\hspace{0.4cm};\hspace{0.6cm} n=0,1,2,3, \dots
\]
and the $a_n(\bar{{\cal K}})$ are the integrals of the functions
defined in Appendix for $\bar{V}(x)=2 \sech^2 x$.

In any case we see from the formula (\ref{asy}) that the
comparison with the exact result is satisfactory:
\[
\Delta M_K\cong -0.282095 \hbar m -\sum_{n=2}^{n_0-1}
a_n(\bar{\cal K}) d_n \hbar m \mbox{ versus } \Delta M_K=-0.318309
\hbar m\,.
\]
The partial sums
\[
D_{n_0}=-\sum_{n=2}^{n_0-1} a_n(\bar{\cal K}) d_n=-\sum_{n=2}^{n_0-1}
a_n(\bar{\cal K}) \frac{\gamma[n-1,1]}{8\pi}
\]
can be estimated with the help of the following Table,
\begin{center}
\begin{tabular}{|c|c|c|c|}
\\[-0.6cm] \hline
$n$ & $a_n(\bar{\cal K})$ & $n_0-1$ & $D_{n_0} $ \\ \hline 2 & 2.66667
& 2 & -0.0670702
\\ 3 & 1.06667 & 3 & -0.0782849\\ 4 & 0.30476 & 4 & -0.0802324
\\ 5 & 0.06772 & 5 & -0.0805373  \\ 6 & 0.012324 & 6 &-0.0805803  \\ 7 &
0.0018944 & 7 &-0.0805857  \\ 8 & 0.00025258 & 8 & -0.0805863 \\ 9
& 0.00002972 & 9 & -0.0805863\\ \hline
\end{tabular}
\end{center}
For instance, choosing $n_0=10$ we find that $D_{10}=-0.080586$
and the correction obtained by means of the asymptotic expansion
is:
\[
E[\psi_K]+\Delta M_K\cong \frac{8m^3}{\lambda} - 0.362681 \hbar
m+o(\hbar^2\lambda)
\]
In fact
\[
\frac{\hbar m}{2}[B_{P\bar{\cal K}}
(-{\textstyle\frac{1}{2}})-B_{\bar{\cal
V}}(-{\textstyle\frac{1}{2}})]+\lim_{L\rightarrow\infty}\frac{\hbar
m}{L} B_{\bar{\cal V}}({\textstyle\frac{1}{2}}) =\frac{\hbar
m}{2\sqrt{\pi}} \int_1^\infty d\beta \left( \frac{1}{2} \frac{{\rm
Erfc \sqrt{\beta}}}{\beta^{\frac{3}{2}}}+\frac{e^{-\beta}}{\beta
\sqrt{\pi}} \right)=0.044373 \hbar m
\]
is almost the total error: $0.044372 \hbar m$. The difference is:
\[
-\frac{\hbar m}{4 \sqrt{\pi}}\,  b_{10,{\cal K}}(-\textstyle\frac{1}{2})
\approx 10^{-6} \hbar m\,.
\]
Note in the Table that $a_n(\bar{\cal K})$ rapidly decreases with
increasing $n$.

\subsection{The quantum $\lambda (\phi^4)_2$ kink}

We now consider the other prototype of solitary waves in
relativistic (1+1)-dimensional field theory: the kink of the
$\lambda (\phi^4)_2$ model. The potential energy density is:
$U[\psi(y^\mu)]= \frac{\lambda}{4}\left(\psi^2-\frac{m^2}{\lambda}
\right)^2$. We choose, however, $m_d=\frac{m}{\sqrt{2}}$ but keep
$c_d=\frac{\sqrt{\lambda}}{m}$ and find:
$\bar{U}[\phi(x,t)]=\frac{1}{2}\left(\phi^2-1\right)^2$.

The internal symmetry group is now the ${\Bbb Z}_2$ group
generated by the  $\phi\rightarrow -\phi$ reflections and the
orbit of vacuum classical configurations $\phi_V(x,t)=\pm 1\in
{\bf M}$ gives rise to a moduli space of vacua $\hat{\bf
M}=\frac{{\bf M}}{{\Bbb Z}_2}$ which is a single point. The kink
solitary waves are thus loop kinks and read
\[
\phi_{\rm K}(x,t)=\pm \tanh x \hspace{1cm},\hspace{1cm} \psi_{\rm K}(y,y_0)=\pm \frac{m}{\sqrt{\lambda}}\, \tanh \frac{m y}{\sqrt{2}}\,.
\]
The kink and vacuum solutions have classical energies of
$E[\psi_{\rm K}]=\frac{4}{3} \frac{m^3}{\sqrt{2}\lambda}$ and
$E[\psi_{\rm V}]=0$ respectively. The Hessian operators for the
vacuum and kink solutions are
\begin{eqnarray*}
{\cal V}&=&-\frac{d^2}{d y^2}+2 m^2=\frac{m^2}{2} \left(
-\frac{d^2}{d x^2}+4 \right)=\frac{m^2}{2} \bar{\cal V} \\ {\cal
K}&=&-\frac{d^2}{d y^2}+2 m^2 -\frac{3 m^2}{{\rm cosh}^2 \frac{m
y}{\sqrt{2}}}=\frac{m^2}{2} \left( -\frac{d^2}{d x^2}+4
-\frac{6}{{\rm cosh}^2 x} \right)=\frac{m^2}{2} \bar{\cal K}
\end{eqnarray*}
and the corresponding generalized zeta functions satisfy
\[
\zeta_{P{\cal K}}(s)=\left( \frac{\sqrt{2}}{m}\right)^{2s} \zeta_{P\bar{\cal K}}(s) \hspace{1cm}, \hspace{1cm} \zeta_{\cal V}(s)=\left( \frac{\sqrt{2}}{m} \right)^{2s} \zeta_{\bar{\cal V}}(s)\,.
\]

\subsubsection{Exact computation of the semi-classical mass and wave
functional}

$\bullet$ Generalized zeta function of $\bar{\cal V}=-\frac{d^2}{d
x^2}+4$.

Acting on the $L^2({\Bbb R})\otimes {\Bbb C}$ Hilbert space we
have that ${\rm Spec}\, \bar{\cal V} =\{k^2+4\}$, $k\in{\Bbb R}$,
but the spectral density on the interval
$\bar{I}=[-\frac{mL}{2\sqrt{2}},\frac{mL}{2\sqrt{2}}]$ of
eigen-functions with periodic boundary conditions is
$\rho_{\bar{\cal V}}(k)=\frac{m L}{2 \sqrt{2}\pi}$. From these
data, the heat and generalized zeta functions are easily computed:
\[
{\rm Tr}\, e^{-\beta \bar{\cal V}}= \frac{mL}{2 \sqrt{2} \pi}
\int_{-\infty}^\infty \hspace{-0.2cm} dk\,
e^{-\beta(k^2+4)}=\frac{mL}{\sqrt{8\pi \beta}}\, e^{-4 \beta}
\]
\begin{equation}
\zeta_{\bar{\cal V}}(s) = \frac{m L}{\sqrt{8 \pi}}
\frac{1}{\Gamma(s)} \int_0^\infty d\beta \, \beta^{s-\frac{3}{2}}
e^{-4 \beta}=\frac{m L}{\sqrt{8 \pi}} \frac{1}{4^{s-\frac{1}{2}}}
\frac{\Gamma(s-\frac{1}{2})}{\Gamma(s)} \label{eq:zetv}
\end{equation}
and we find that $\zeta_{\bar{\cal V}}(s)$ has the same poles and
infrared behaviour in the $\lambda (\phi^4 )_2$ and the
sine-Gordon models.

\vspace{0.2cm}

\noindent $\bullet$ Generalized zeta function of $\bar{\cal
K}=-\frac{d^2}{d x^2}+4-\frac{6}{\cosh^2 x}$.

$\bar{\cal K}$ is the Schr\"odinger operator for the second
transparent Posch-Teller potential, \cite{Drazin}. Thus, ${\rm
Spec}\, \bar{\cal K}=\{0\}\cup\{3\}\cup \{k^2+4\}$, $k\in {\Bbb
R}$, and the spectral density on $I$ is
\[
\rho_{\bar{\cal K}}(k)=\frac{mL}{2\sqrt{2}\pi}+\frac{1}{2 \pi}
\frac{d \delta(k)}{d k}
\]
where the phase shifts are $\delta(k)=-2 \arctan
\frac{3k}{2-k^2}$, if PBC are considered. Thus, we find
\begin{eqnarray*}
{\rm Tr} e^{-\beta\,P \bar{\cal K}}&=& e^{-3\beta}+\frac{mL}{\sqrt{8} \pi} \int_{-\infty}^\infty \hspace{-0.2cm} dk \, e^{-\beta (k^2+4)}+\frac{1}{2\pi} \int_{-\infty}^\infty \hspace{-0.2cm} dk \, \frac{d \delta(k)}{d k} e^{-\beta(k^2+4)} \\
&=& \frac{m L}{\sqrt{8\pi \beta}}\, e^{-4 \beta} +e^{-3 \beta}
(1-{\rm Erfc \sqrt{\beta}})-{\rm Erfc}\, 2\sqrt{\beta}\,.
\end{eqnarray*}
The Mellin transform immediately provides the
generalized zeta function:
\begin{equation}
{\zeta_{P\bar{\cal K}}}(s)=\zeta_{\bar{\cal V}}(s)+
\frac{\Gamma(s+\frac{1}{2})}{\sqrt{\pi}\,\Gamma(s)} \left[
\frac{2}{3^{s+\frac{1}{2}}} {}_2F_{1}
[{\textstyle\frac{1}{2}},{s+\textstyle\frac{1}{2}},{\textstyle\frac{3}{2}},
-{\textstyle\frac{1}{3}}]-\frac{1}{4^s} \frac{1}{s} \right]
\label{zetk}
\end{equation}
where ${}_2 F_1[a,b,c,d]$ is the Gauss hypergeometric function,
\cite{Abramowitz}.

The power expansion of ${}_2 F_1$,
\[
{}_2
F_{1}[{\textstyle\frac{1}{2}},{s+\textstyle\frac{1}{2}},{\textstyle\frac{3}{2}},
-{\textstyle\frac{1}{3}}]=\frac{\Gamma (\frac{3}{2})}{\Gamma
(\frac{1}{2})\Gamma (s+\frac{1}{2})}\sum_{l=0}^\infty
\frac{(-1)^l}{3^l l!}\frac{\Gamma (l+\frac{1}{2})\Gamma
(s+l+\frac{1}{2})}{\Gamma (l+\frac{3}{2})}
\]
tells us that, besides the poles of $\zeta_{\bar{\cal V}}(s)$,
${\zeta_{P\bar{\cal K}}}(s)$ has poles at
$s=-\frac{1}{2}+l,-\frac{3}{2}+l,-\frac{5}{2}+l,\cdots$, $l\in
{\Bbb Z}^{+} \cup \{0\}$; i.e., as in the sG soliton case,
$\zeta_{\bar{\cal V}}(s)$ and ${\zeta_{P\bar{\cal K}}}(s)$ share
the same poles except $s=\frac{1}{2}$ but the residues in the
$\lambda (\phi^4)_2$ model are increasingly different with larger
and larger values of $|{\rm Re}\hspace{0.1cm}{
 s}|$.

Applying these results to the kink mass formula, we obtain
\begin{eqnarray*}
\Delta_1\varepsilon^{\rm K}&=&\lim_{s \rightarrow -\frac{1}{2}}\frac{\hbar}{2}
\left( \frac{2 \mu^2}{m^2} \right)^s \mu \left[ \zeta_{P\bar{\cal
K}} (s)-\zeta_{\bar{\cal V}}(s) \right]\\&=& \frac{\hbar m}{2\sqrt{2\pi}}
\lim_{\varepsilon \rightarrow 0}\left( \frac{2 \mu^2}{m^2}
\right)^{\varepsilon}
\frac{\Gamma(\varepsilon)}{\Gamma(-\frac{1}{2}+\varepsilon)}
\left[ \frac{2}{3^\varepsilon} \, {}_2
F_1[{\textstyle\frac{1}{2}},\varepsilon,
{\textstyle\frac{3}{2}},-{\textstyle\frac{1}{3}}]-
\frac{1}{(-\frac{1}{2}+\varepsilon)\,4^{-\frac{1}{2}+\varepsilon}}
\right]\\ &=& \frac{\hbar m}{2\sqrt{2}\pi} \lim_{\varepsilon
\rightarrow 0} \left[ -\frac{3}{\varepsilon}-3
\ln \frac{2 \mu^2}{m^2}+2 +\ln \frac{3}{4}
-{}_2F_1'[{\textstyle\frac{1}{2}},0,{\textstyle\frac{3}{2}},-{\textstyle\frac{1}{3}}]
+o(\varepsilon) \right]
\end{eqnarray*}
\begin{eqnarray*}
\Delta_2 \varepsilon^{\rm
K}&=&-\lim_{L\rightarrow\infty}\frac{6\hbar}{L} \lim_{s\rightarrow
-\frac{1}{2}} \left( \frac{2 \mu^2}{m^2}
\right)^{s+\frac{1}{2}}\frac{\Gamma[s+1]}{\Gamma[s]}
\zeta_{\bar{\cal V}}(s+1)=-\frac{3 \hbar m}{\sqrt{2\pi}}
\lim_{\varepsilon \rightarrow 0} \left( \frac{2 \mu^2}{m^2}
\right)^\varepsilon \frac{4^{-\varepsilon}
\Gamma(\varepsilon)}{\Gamma(-\frac{1}{2}+\varepsilon)}\\&+&o(\hbar^2\lambda)=\frac{3\hbar
m}{2 \sqrt{2}\pi} \lim_{\varepsilon \rightarrow 0} \left[
\frac{1}{\varepsilon}+ \ln \frac{2 \mu^2}{m^2}-\ln 4+\psi(1)-\psi(-\frac{1}{2})+o(\varepsilon)
\right]+o(\hbar^2 \lambda)
\end{eqnarray*}
where ${}_2 F_1'$ is the derivative of the Gauss hypergeometric
function with respect to the second argument. Therefore,
$\Delta_1\varepsilon^{\rm K}+\Delta_2\varepsilon^{\rm
K}=\frac{\hbar m}{2\sqrt{6}}-\frac{3\hbar m}{\pi\sqrt{2}}\,$, and
we obtain:
\[
\varepsilon_R^{\rm K}=E[\psi_{\rm K}]+\Delta M_{\rm K}=\frac{4}{3}
\frac{m^3}{\sqrt{2}\lambda}+ \hbar
m\left(\frac{1}{2\sqrt{6}}-\frac{3}{\pi\sqrt{2}}\right)+o(\hbar^2\lambda)\quad,
\]
the same answer as offered by the mode-number regularization
method \cite{Rebhan}.

To compute the norm of the ground state wave functionals we
closely follow the procedure applied in sub-Section \S 3.1. to the
sine-Gordon soliton. In the $\lambda(\phi^4)_2$ model, we find
that
\begin{eqnarray*}
\frac{d \zeta_{P\bar{\cal K}}}{ds}&=& \frac{d \zeta_{\bar{\cal
V}}}{ds}+\frac{1}{\sqrt{\pi}}
\frac{\Gamma(s+\frac{1}{2})}{\Gamma(s)}\left[ 4^{-s} \left(
\frac{1}{s}+\ln 4 +\psi(s)-\psi(s+{\textstyle\frac{1}{2}}) \right)
\right.-\\
&& \hspace{-2cm} \left. - 2 s \, 3^{-s-\frac{1}{2}} \, {}_2F_{1}
[{\textstyle\frac{1}{2}},{s+\textstyle\frac{1}{2}},
{\textstyle\frac{3}{2}},-{\textstyle\frac{1}{3}}] ( \log 3-
\psi(s+{\textstyle\frac{1}{2}})+\psi(s))+2 s \, 3^{-s-\frac{1}{2}}
{}_2F_{1}' [{\textstyle\frac{1}{2}},{s+\textstyle\frac{1}{2}},
{\textstyle\frac{3}{2}},-{\textstyle\frac{1}{3}}]\right]
\end{eqnarray*}
and
\[
\frac{d \zeta_{\bar{\cal
V}}}{ds}=\frac{mL}{\sqrt{8\pi}}\frac{1}{4^{s-\frac{1}{2}}}
\frac{\Gamma(s-\frac{1}{2})}{\Gamma(s)}\,
\left(\psi(s-{\textstyle\frac{1}{2}})-\psi(s)-\log 4 \right)\,.
\]

from these expressions and formulas (\ref{eq:zetv}) and
(\ref{zetk}) one checks that
\[
\zeta_{\bar{\cal V}}(0)=0 \hspace{0.2cm}, \hspace{0.2cm}
\zeta_{P\bar{\cal K}}(0)=-1\hspace{0.2cm}, \hspace{0.2cm}
\frac{d \zeta_{\bar{\cal V}}}{ds}(0)= -\sqrt{2}mL\hspace{0.2cm},
\hspace{0.2cm} \frac{d \zeta_{P\bar{\cal K}}}{ds}(0)=
-\sqrt{2}mL+\log 48\, .
\]
We obtain
\begin{displaymath}
\left|
\Psi_0(\phi_{\rm K}(x))\right|^2 =\frac{1}{2}\left
(\frac{C}{\sqrt{3}}\right )^{\frac{1}{2}}\, \exp \left [ \frac{m
L}{2\sqrt{2}}\right ],\ \ \ \ \ \
\left|
\Psi_0(\phi_{\rm V}(x))\right|^2 = \exp \left [\frac{m
L}{2\sqrt{2}}\right ]\,.
\end{displaymath}
The quotient of the probability densities is
\begin{equation}
\frac{\left| \Psi_0(\phi_{\rm K}(x))\right|^2}{
\left| \Psi_0(\phi_{\rm V}(x))\right|^2} = \frac{1}{2}\left
(\frac{C}{\sqrt{3}}\right )^{\frac{1}{2}}\,. \label{eq:div2}
\end{equation}

\subsubsection{The asymptotic expansion and quantum corrections}

In the $\lambda (\phi^4)_2$ model
$\frac{d^2\bar{U}}{d\phi^2}|_{\phi_{\rm V}}(x)=4$ and
$\bar{V}(x)=\frac{d^2\bar{U }}{d\phi^2}|_{\phi_{\rm
V}}(x)-\frac{d^2\bar{U}}{d\phi^2}|_{\phi_{\rm K}}(x)=6 \sech^2 x$
are the potentials of the Sch\"odinger operators that
respectively correspond to the Hessians at the vacuum and the
kink configurations. The asymptotic expansion of the heat function
\[
{\rm Tr} e^{-\beta\,P \bar{\cal K}}=-1+\frac{e^{-4
\beta}}{\sqrt{4\pi \beta}} \sum_{n=0}^\infty
\int_{-\frac{mL}{2\sqrt{2}}}^\frac{mL}{2\sqrt{2}}\hspace{-0.3cm}
dx \, a_n(x,x) \beta^n=-1+\frac{e^{-4 \beta}}{\sqrt{4 \pi}}
\sum_{n=0}^\infty a_n(\bar{\cal K})\, \beta^{n-\frac{1}{2}}
\]
can be either obtained as a series expansion of the exact result
\[
{\rm Tr} e^{-\beta\,P \bar{\cal K}}=-1+\left[
\frac{mL}{\sqrt{8\pi \beta}}+\frac{1}{\sqrt{\pi}}
\sum_{n=1}^\infty \frac{2^{n}(1+2^{2n-1})}{(2n-1)!!}
\beta^{n-\frac{1}{2}}\right]e^{-4 \beta}
\]
or from the coefficients defined in the Appendix for $\bar{V}(x)=6
\sech^2 x$
\[
a_n(\bar{\cal
K})=\int_{-\frac{mL}{2\sqrt{2}}}^{\frac{mL}{2\sqrt{2}}}
\hspace{-0.3cm} dx \, a_{n}(x,x)\hspace{0.5cm},\hspace{0.5cm}
a_0(\bar{\cal K})=\frac{mL}{\sqrt{2}}\hspace{0.5cm},\hspace{0.5cm}
a_{n}(\bar{\cal K})=\frac{2^{n+1}(1+2^{2n-1})}{(2 n-1)!!}\,.
\]

To compare with the exact result, we apply the formula given in
the Appendix and observe that
\[
\varepsilon_R^{\rm K} \cong \frac{4}{3} \frac{m^3}{\sqrt{2}
\lambda} -0.199471 \hbar m-\sum_{n=2}^{n_0-1} a_n(\bar{\cal K})
d_n \hbar m + o(\hbar^2\lambda)
\]
is far from the exact result
\[
\varepsilon_R^{\rm K} = \frac{4}{3} \frac{m^3}{\sqrt{2}
\lambda}-0.471113 \hbar m +o(\hbar^2\lambda)
\]
before adding the contribution of the terms between $n=2$ and
$n=n_0-1$ in the asymptotic expansion to the contribution coming
from the subtraction  of the translational mode. The partial sums
\[
D_{n_0}=-\sum_{n=2}^{n_0-1} a_n(\bar{\cal K}) d_n =- \sum_{n=2}^{n_0-1}
a_n(\bar{\cal K})\, \frac{\gamma[n-1,4]}{8 \sqrt{2} \pi\, 4^{n-1} }
\]
can be estimated up to $n_0=11$ with the help of the following
Table {\small
\begin{center}
\begin{tabular}{|c|c|c|c|}
\\[-0.5cm] \hline
$n$ & $a_n (\bar{\cal K})$ & $n_0-1$ & $D_{n_0}$ \\ \hline 2 & 24.0000
& 2 & -0.165717 \\ 3 & 35.2000 & 3 &-0.221946 \\ 4 & 39.3143 & 4
&-0.248281  \\ 5 & 34.7429 & 5 &-0.261260  \\ 6 & 25.2306 & 6 &
-0.267436  \\ 7 & 15.5208 & 7 & -0.270186  \\ 8 & 8.27702 & 8 &
-0.271317 \\ 9 & 3.89498 & 9 & -0.271748 \\ 10 & 1.63998 & 10
&-0.271900 \\ \hline
\end{tabular}
\end{center}}

Choosing $n_0=11$,we find that $D_{11}=-0.271900 \hbar m$ and the
correction obtained by adding ${\rm D}_{11}\hbar m$ is:
\[
\Delta M_{\rm K} \cong -0.471371 \hbar m+o(\hbar^2\lambda)
\]
in good agreement with the exact result above. In fact
\begin{eqnarray*}
&& \frac{\hbar m}{2}[B_{P\bar{\cal K}}
(-{\textstyle\frac{1}{2}})-B_{\bar{\cal V}}(-{\textstyle\frac{1}{2}})]+\frac{3 \hbar m}{\sqrt{2} L} B_{\bar{\cal V}}({\textstyle\frac{1}{2}}) \\ &&=\frac{\hbar
m}{2 \sqrt{2\pi}} \int_1^\infty d\beta \left( -\frac{e^{-3
\beta}}{2 \beta^{\frac{3}{2}}}+\frac{e^{-3\beta} {\rm Erfc
\sqrt{\beta}}}{2 \beta^{\frac{3}{2}}}+\frac{{\rm
Erfc}\,2\sqrt{\beta}}{2\beta^{\frac{3}{2}}}+\frac{3e^{-4\beta}}{\sqrt{\pi}\beta}
\right) \approx  0.00032792 \hbar m
\end{eqnarray*}
is almost the total error: $0.0002580 \hbar m$. The deviation is

\[
-\frac{\hbar m}{4\sqrt{2\pi}}b_{10,{\cal K}}(-\frac{1}{2})
 \approx 10^{-4} \hbar m\,.
\]

With respect to the sine-Gordon model there are two differences:
a) in the $\lambda(\phi^4)_2$ model the error committed by using
asymptotic methods is smaller, of the order of $10^{-4} \hbar m$,
a 0.07 percent, as compared with $10^{-2} \hbar m$, a 6.00
percent, in the sG case; b) the rejection of the contributions of
the $n_0>11$ terms and the non-exact computation of the mass
counter-term contribution has a cost of approximately $10^{-4}
\hbar m$ in the $\lambda(\phi^4)_2$ model versus $10^{-6} \hbar m$
in the sG system. Both facts have to do with the larger value of
the smaller eigenvalue of the vacuum Hessian in the $\lambda
(\phi^4)_2$ model with respect to the sG system, 4 versus 1.

\subsection{The cubic sinh-Gordon kink}

We shall now study a system of the same type where the potential
energy density is: $ U[\psi(y^\mu)]=\frac{m^4}{4\lambda}
\left({\rm sinh}^2\frac{\sqrt{\lambda}}{m}\psi-1 \right)^2$. Non-dimensional quantities are defined through the choice $m_d=m$
and $c_d=\frac{\sqrt{\lambda}}{m}$; the Euler-Lagrange equation is
\begin{equation}
\Box \phi(t,x)=-\frac{1}{2}{\rm sinh}(2\phi)({\sinh^2}\phi-1)
\end{equation}
and the justification for the choice of name is clear. We find
this model interesting because it reduces to the
$\lambda(\phi^4)_2$ system if $|\phi(t,x)|<1$ and is the Liouville
model, \cite{Liouv}, with opposite sign of the coupling constant,
in the $\phi(t,x)\cong \pm\infty$ ranges. In fact, the potential
energy density $\bar{U}(\phi)=\frac{1}{4}(\sinh^2 \phi -1  )^2$,
see Figure 1(a), presents two minima at the classical values:
$\phi_{\rm V}=\pm {\rm arcsinh}1$. The two vacuum points are
identified by the $\phi\rightarrow -\phi$ internal symmetry
transformation and the semi-classical vacuum moduli space is a
point. For this reason, $\bar{U}(x)$ has been applied to the study
of the quantum theory of diatomic molecules: the solutions of the
associated time-independent Schr\"odinger equation are a good
approximation to the eigen-states of a quantum particle that moves
under the influence of two centers of force. We deal with the
$\zeta=1$ and $M=3$ member of the Razavy family of
quasi-exactly-solvable Schr\"odinger operators, \cite{Razavy},
although we are looking at it from a field-theoretical
perspective.

The solutions of the first-order equations
\begin{equation}
\frac{d\phi}{dx}=\pm\frac{1}{\sqrt{2}}({\rm
sinh}^2\phi-1)\hspace{0.5cm},\hspace{0.3cm}\phi_{\rm K}(x)= \pm \,
{\rm arctanh} \, \frac{\tanh (x+b)}{\sqrt{2}}  ,
\end{equation}
see Figure 1(b) for $b=0$, are the kink solitary waves of the
system.
\begin{figure}[htbp]
\centerline{\psfig{file=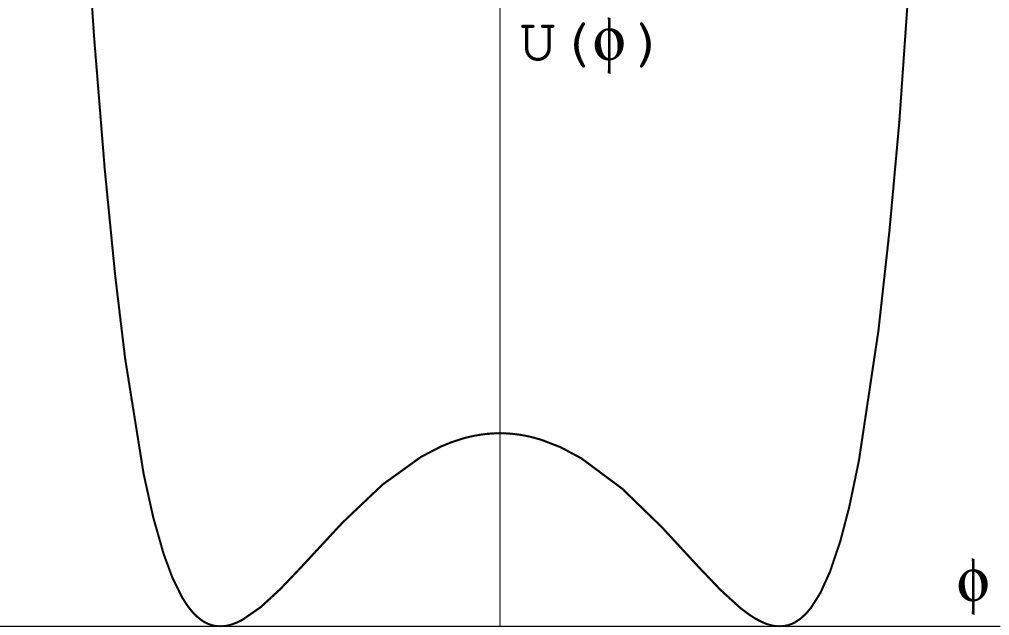, height=3.cm} \hspace{0.6cm}
\psfig{file=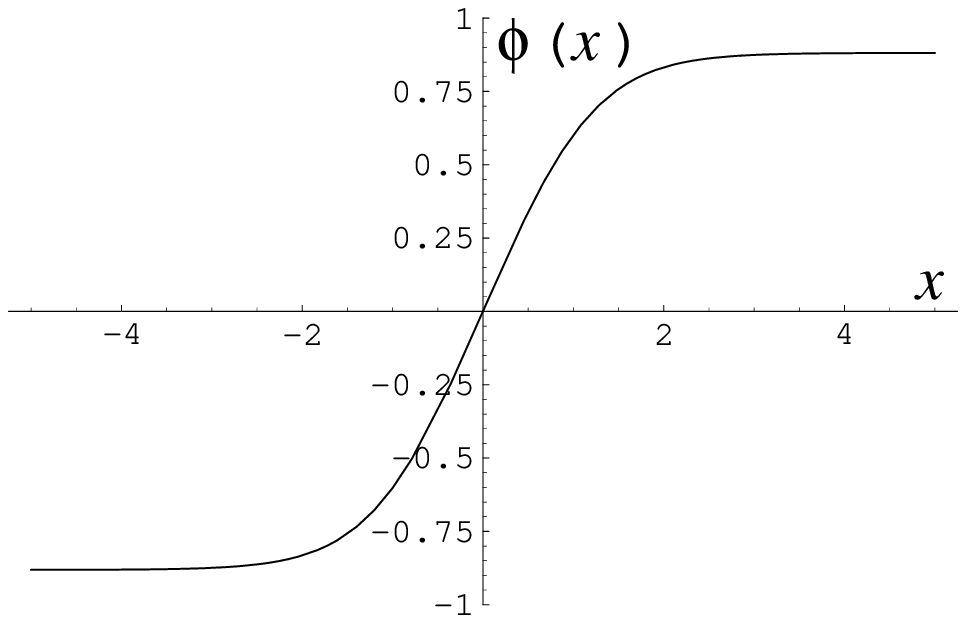,height=3.cm}\hspace{0.6cm}
\psfig{file=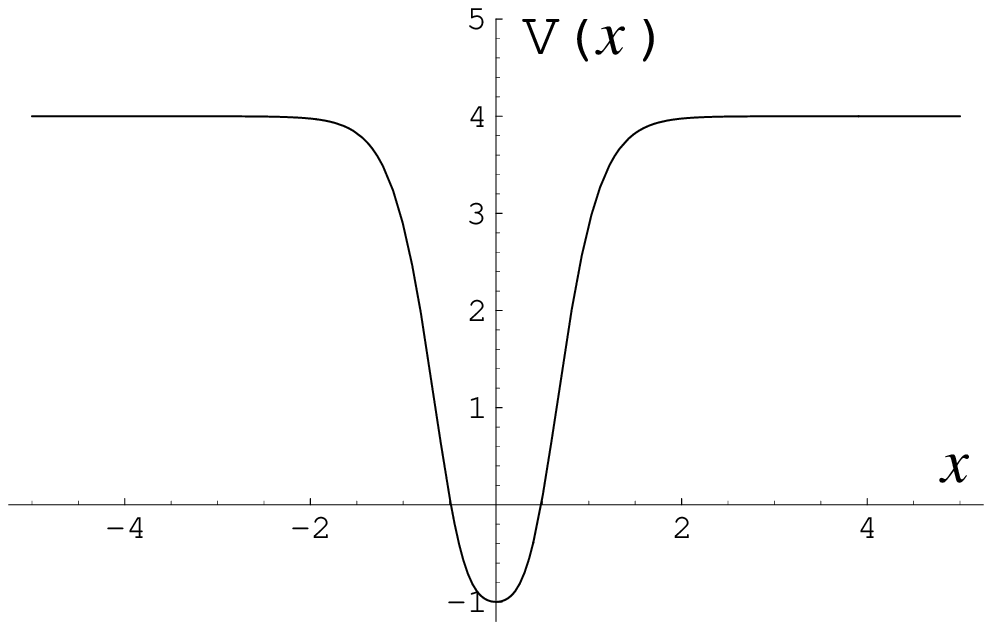,height=3.cm} } \caption{\small \it Graphic
representation of (a) the potential energy density (b) the kink
and (c) the Hessian potential well.}
\end{figure}
The Hessian operators at the vacuum and kink solutions are
respectively:
\[
{\cal
V}=-\frac{d^2}{dy^2}+4m^2=m^2\left(-\frac{d^2}{dx^2}+4\right)=m^2
\, \bar{\cal V}
\]
\[
{\cal K}=-\frac{d^2}{dy^2}+2m^2+\frac{16m^2}{(1+\sech^2
y)^2}-\frac{14m^2}{1+\sech^2
y}=m^2(-\frac{d^2}{dx^2}+2+\frac{16}{(1+\sech^2
x)^2}-\frac{14}{1+\sech^2 x})=m^2\bar{\cal K}
\]
The mass of the fundamental mesons is thus $2m$.  $\bar{\cal K}$
is an Schr\"odinger operator:
\[
\bar{\cal
K}=-\frac{d^2}{dx^2}+4-\bar{V}(x)\hspace{0.6cm},\hspace{0.4cm}\bar{V}(x)=\frac{2\,{\rm
sech}^2x(9+{\rm sech}^2x)}{(1+{\rm sech}^2x)^2}
\]
where the potential well plotted in Figure 1(c), albeit
analytically very different from the s-G and $\lambda (\phi^4)_2$
kink potential wells, exhibits a similar shape.

We shall not attempt to solve the spectral problem of $\bar{\cal
K}$. The only thing that we need to know in order to apply the
asymptotic method is that the lowest eigen-state is the unique
zero mode:
\[
f_0(x)= \frac{d \phi_{\rm K}}{d x}=\frac{2\sqrt{2}}{(3+\cosh 2 x)}
\]
Therefore, the energy of the semi-classical kink state is
approximately (see formula (\ref{asy}))
\begin{equation}
\varepsilon_R^{\rm K} \cong
\frac{m^3}{2\lambda}\left(1-3\sqrt{2}{\rm arcsinh}1\right)-\hbar
m\left(\frac{1}{2\sqrt{\pi}}+\sum_{n=2}^{n_0-1}a_n(\bar{\cal
K})\frac{\gamma [n-1,4]}{8\pi 4^{n-1}}\right)
\end{equation}
In the Table below we write the Seeley's coefficients and the
partial sums $D_{n_0}=-\sum_{n=2}^{n_0-1}a_n(\bar{\cal
K})\frac{\gamma [n-1,4]}{8\pi 4^{n-1}}$ up to $n_0=11$:

\begin{center}
\begin{tabular}{|c|c|c|c|}
\\[-0.5cm] \hline
$n$ & $a_n (\bar{\cal K})$ & $n_0-1$ & $D_{n_0}$ \\ \hline 2 &
29.1604 & 2 & -0.20135 \\ 3 & 39.8523 & 3 &-0.26501 \\ 4 & 42.1618
& 4 &-0.293253  \\ 5 & 36.0361 & 5 &-0.306715  \\ 6 & 25.7003 & 6
& -0.313005  \\ 7 & 15.6633 & 7 & -0.315779  \\ 8 & 8.3143 & 8 &
-0.316917 \\ 9 & 3.9033 & 9 & -0.317349 \\ 10 & 1.6590 & 10
&-0.317502 \\ \hline
\end{tabular}
\end{center}
obtaining the approximate answer:
\[
\Delta M_{\rm K} \cong \hbar m[\Delta_0+ D_{11}]=-0.282095 \hbar m
-0.317502 \hbar m = -0.599597 \hbar m
\]
We cannot estimate the error but we assume that this result is as
good as the answer obtained for the $\lambda(\phi^4)_2$ kink
because the continuous spectrum of $\bar{\cal K}$ also starts at
4.

\section{Link kinks: the $\lambda(\phi^6)_2$ model}

Finally, we consider the following potential energy density:
 $ U[\psi(y^\mu)]=\frac{\lambda^2}{4m^2}\psi^2
\left(\psi^2-\frac{m^2}{\lambda} \right)^2$. The choice of
$m_d=\frac{m}{\sqrt{2}}$ and $c_d=\frac{\sqrt{\lambda}}{m}$ leads
to the non-dimensional potential: $
\bar{U}[\phi(x,t)]=\frac{1}{2}\phi^2 \left(\phi^2-1\right)^2$. The
moduli space of vacua $\hat{\bf M}=\frac{{\bf M}}{{\Bbb Z}_2}$,
made out of two ${\Bbb Z}_2$ orbits, contains two points:
\[
\phi_{V_0}(x,t)=0
\hspace{0.5cm},\hspace{0.5cm}\phi_{V_\pm}(x,t)=\pm 1 .
\]
Quantization around the $\phi_{V_0}(x,t)$ vacuum preserves the
$\phi\rightarrow -\phi$ symmetry, which is spontaneously broken at
the degenerate vacua $\phi_{V_\pm}(x,t)$. The kink solitary waves
of the system
\[
\phi_{\rm K}(x,t)=\pm\frac{1}{\sqrt{2}}\sqrt{1\pm{\rm
tanh}(x+b)}\hspace{0.5cm},\hspace{0.5cm}\psi_{\rm K}
(y,y^0)=\pm\frac{m}{\sqrt{2\lambda}}\sqrt{1\pm{\rm
tanh}\frac{m}{\sqrt{2}}(y+{\tilde b})}
\]
interpolate between $\phi_{V_\pm}(x,t)$ and $\phi_{V_0}(x,t)$, or
vice-versa, which are vacua belonging to distinct ${\Bbb Z}_2$
orbits: these solutions are thus link kinks.

The kink and vacuum solutions have classical energies of
$E[\psi_{\rm K}]=\frac{1}{4} \frac{m^3}{\sqrt{2}\lambda}$ and
$E[\psi_{V_0}]=E[\psi_{V_\pm}]=0$ respectively. The Hessian
operators for the vacuum and kink solutions are
\begin{eqnarray*}
{\cal V}_0&=&-\frac{d^2}{d y^2}+\frac{m^2}{2}=\frac{m^2}{2} \left(
-\frac{d^2}{d x^2}+1 \right)=\frac{m^2}{2} \bar{\cal V}_0 \\ {\cal
V}_\pm&=&-\frac{d^2}{d y^2}+2 m^2=\frac{m^2}{2} \left(
-\frac{d^2}{d x^2}+4 \right)=\frac{m^2}{2} \bar{\cal V}_\pm
\\{\cal K}&=&-\frac{d^2}{d y^2}+\frac{5m^2}{4}\pm\frac{3m^2}{4}{\rm tanh}\frac{my}{\sqrt{2}} -
\frac{15m^2}{8{\rm cosh}^2 \frac{m
y}{\sqrt{2}}}=\\&=&\frac{m^2}{2} \left( -\frac{d^2}{d
x^2}+\frac{5}{2}\pm\frac{3}{2}{\rm tanh}x -\frac{15}{4{\rm cosh}^2
x} \right)=\frac{m^2}{2} \bar{\cal K}
\end{eqnarray*}
\begin{figure}[htbp]
\centerline{\psfig{file=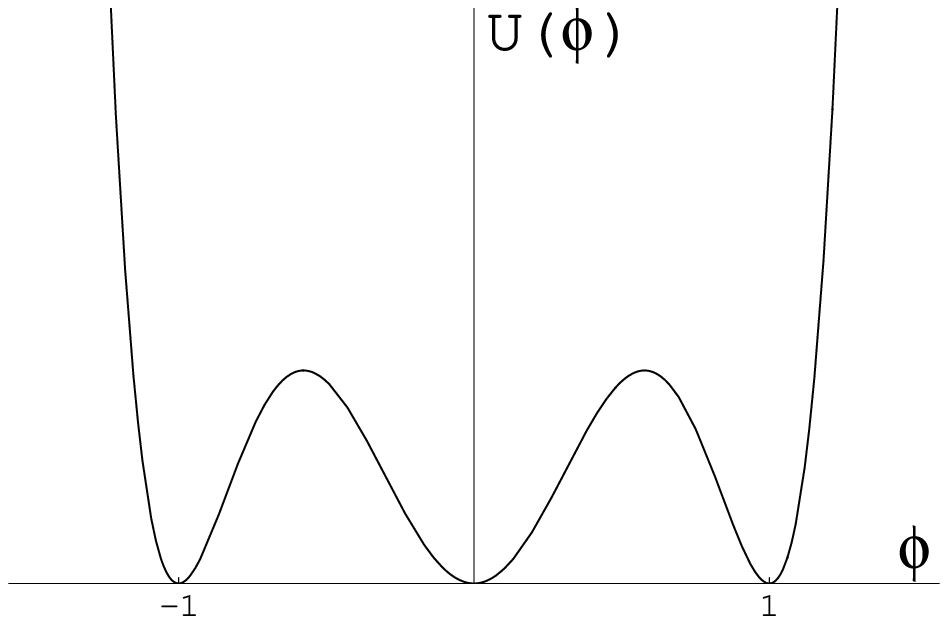, height=3.cm} \hspace{0.6cm}
\psfig{file=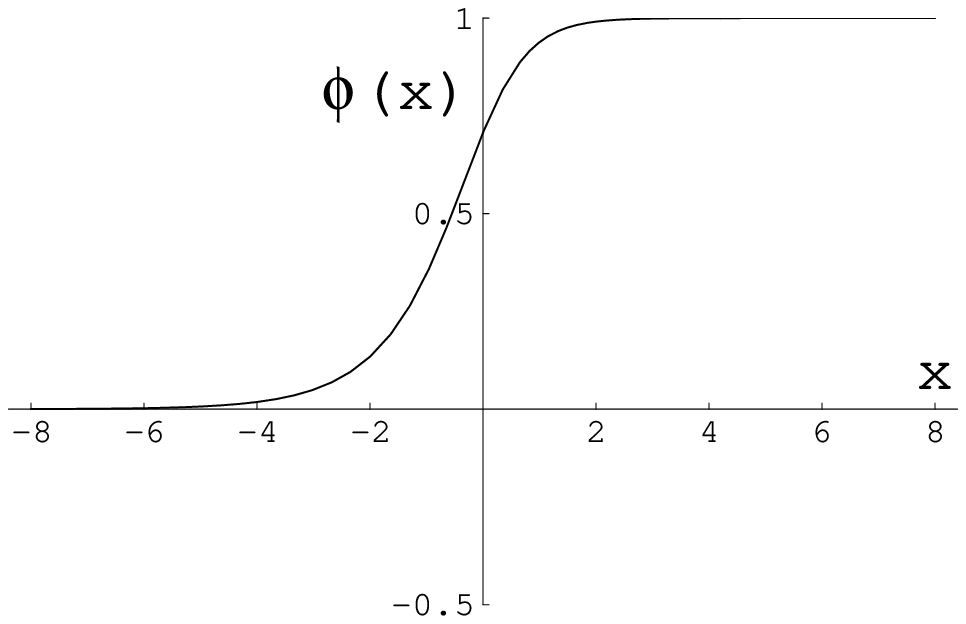,height=3.cm}\hspace{0.6cm}
\psfig{file=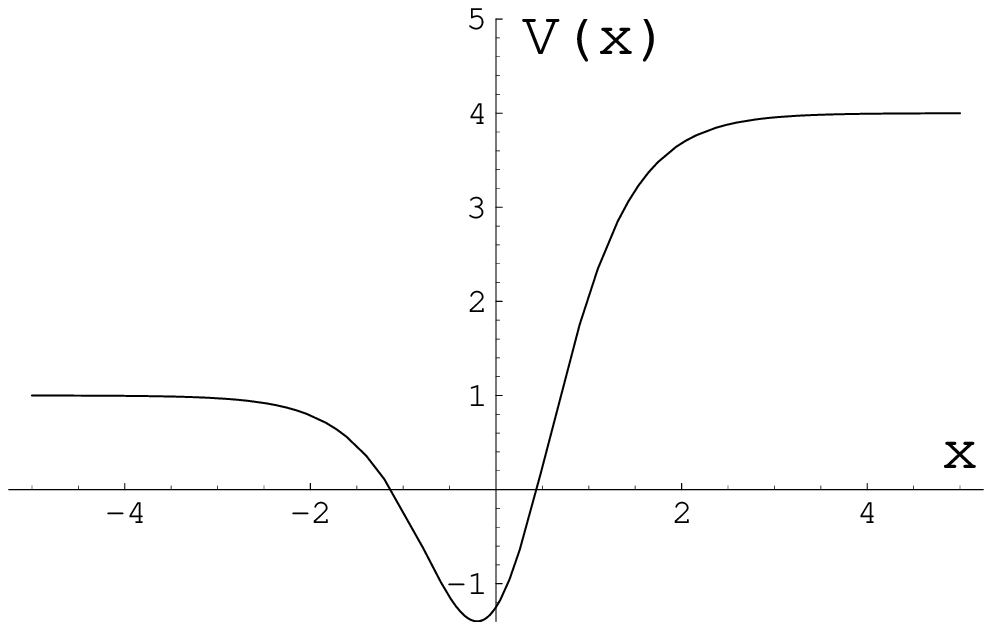,height=3.cm} } \caption{\small \it
Graphical representation of (a) the potential energy density (b)
the kink and (c) the Hessian potential well.}
\end{figure}

The problem of the semi-classical quantization of these and other
link kinks have been addressed somewhat unsuccessfully in
\cite{Lohe} due to the analytical complexity of the
eigen-functions of $\bar{\cal K}$ as well as the conceptual
difficulty of dealing with a QFT on the real line where the
asymptotic states far on the left and far on the right correspond
to mesons with different masses. This issue has been analyzed in
depth in \cite{O'Brien}: the main suggestion is that the
normal-order prescription should be performed with an arbitrary
mass to be fixed in order to avoid the ambiguity induced by the
step function background. We now apply the asymptotic expansion of
the heat function method in this complex circumstance to find a
very natural way of choosing the mass renormalization parameter.
Moreover, we improve the approximation obtained in the computation
of the quantum kink mass by going farther than first-order in the
asymptotic expansion.

Besides the bound state,
\[
f_0(x)=\frac{1}{2{\rm cosh}^2x\sqrt{2(1\pm{\rm tanh}x)}} ,
\]
the $\omega^2=0$ translational mode, the spectrum of $\bar{\cal
K}$ includes transmissionless scattering states for
$1\leq\omega^2\leq 4$, and states with both non null transmission
and reflection coefficients if $\omega^2\geq 4$. In the language
of QFT, the topological sectors based on link kinks are peculiar
in the sense that the $N$-particle asymptotic states are mesons
that have different masses at $x=\pm\infty$. If the meson energy
is less than $2m^2$, the bosons are reflected when coming from the
left/right towards the kink. More energetic mesons can either be
reflected by or pass through the kink. If the mesons are
transmitted there is a conversion from kinetic to \lq\lq inertial"
energy, or vice-versa, in such a way that the poles of the
propagators far to the left or far to the right of the kink can
only occur at $p^2=\frac{m^2}{\sqrt{2}}$ and $p^2=2m^2$.

This is the reason why the subtraction from the Casimir energy of
$\phi_{\rm K}$, $\frac{\hbar}{2}\zeta_{P{\cal K}}(-\frac{1}{2})$,
of either the Casimir energy of the $\phi_{V_0}$,
$\frac{\hbar}{2}\zeta_{{\cal V}_0}(-\frac{1}{2})$, or the
$\phi_{V_\pm}$, $\frac{\hbar}{2}\zeta_{{\cal
V}_\pm}(-\frac{1}{2})$, vacua is hopeless, even after adding the
mass renormalization counter-term to the Lagrangian. Therefore, we
cannot use the generalized zeta functions $ \zeta_{{\cal V}_0}(s)$
and $ \zeta_{{\cal V}_\pm}(s)$ to renormalize the zero point
energy in the kink sector. Instead, we will gauge the kink Casimir
energy against the Casimir energies of a family of background
field configurations that satisfy:
\begin{equation}
5\phi_B^4(x)-4\phi_B^2(x)=\frac{1}{2}(1\pm{\rm tanh}\alpha
x),\label{back}
\end{equation}
where $\alpha\in {\Bbb R}^+$. The rationale behind this choice is
that the $\alpha\rightarrow\infty$ limit is the background used by
Lohe, \cite{Lohe}: $\phi_{\bar{B}_\infty}(x)=\pm\theta(x)$. The
problem with Lohe's choice is that the discontinuity at the origin
poses many problems for the algorithm of the asymptotic expansion
because a nightmare of delta functions and their derivatives
appears at $x=0$ at orders higher than the first. Thus, we need
some regularization, which is achieved by replacing the sign
function by tanh in the formula (\ref{back}) above. In Figures
3(a) and 3(b) the Hessian potential wells for the backgrounds
$\phi_{\bar{B}_\infty}$ and $\phi_{\bar{B}_1}$ are compared.
\begin{figure}[htbp]
\centerline{\psfig{file=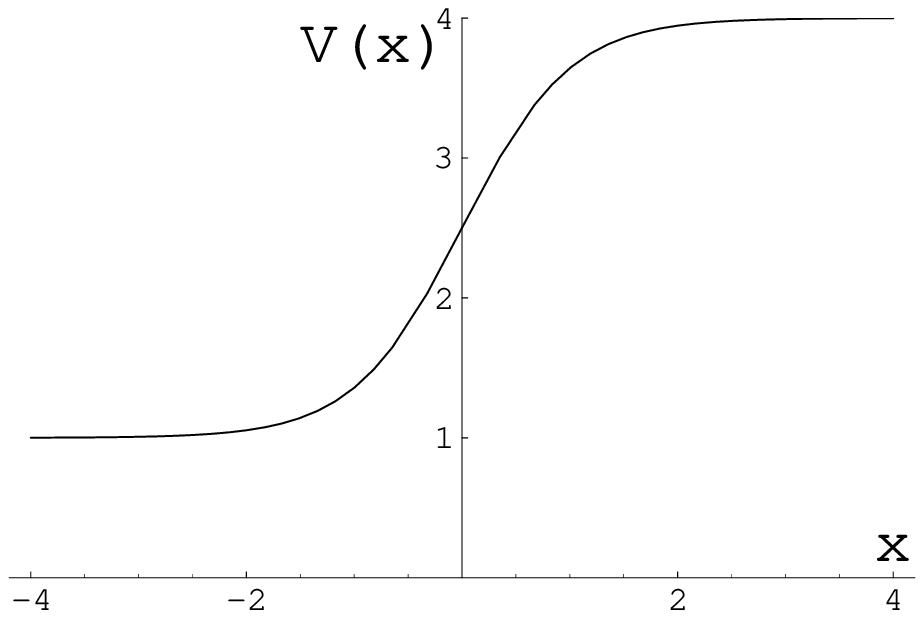, height=3.cm} \hspace{1.8cm}
\psfig{file=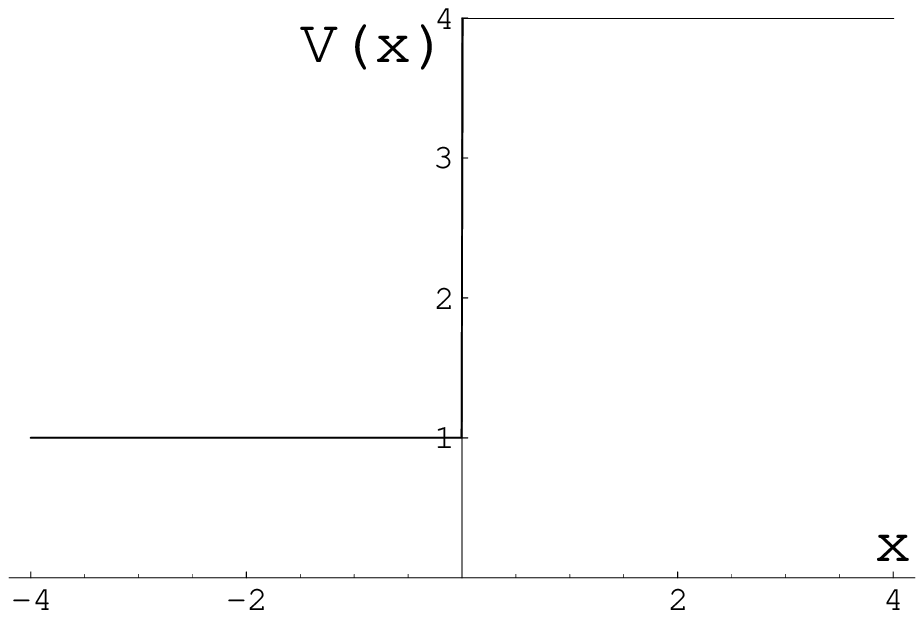,height=3.cm} } \caption{\small \it Graphic
representation of the potential well produced by (a) the
$\phi_{B_1}$ background (b) the $\phi_{B_{\infty}}$ background as
functions of x}
\end{figure}

For any non-zero finite $\alpha$, $\phi_{\bar{B}_\alpha}(x)$
interpolates smoothly between $\pm\frac{4}{5}$ and $\pm 1$ when
$x$ varies from $-\infty$ to $\infty$. The jump from $\pm 1$ to 0
occurring at $x=0$ in $\phi_{\bar{B}_\infty}(x)$ becomes a jump
from $\pm\frac{4}{5}$ to 0, which therefore takes place at
$x=\pm\infty$!, followed by the smooth interpolation to $\pm 1$ .
If $\alpha=0$ the background configuration is also pathological:
$\phi_{\bar{B}_0}(x)=\pm\frac{2+\sqrt{\frac{13}{2}}}{5},\forall
x$, except at $x=\pm\infty$, where there are jumps to 0 and $\pm
1$.

The Schr\"odinger operators
\begin{eqnarray*}
\bar{\cal
B}_\alpha=-\frac{d^2}{dx^2}+\frac{5}{2}&\pm&\frac{3}{2}{\rm
tanh}\alpha x
\\\bar{\cal B}_\infty=-\frac{d^2}{dx^2}+\frac{5}{2}\pm\frac{3}{2}\varepsilon(x)
\hspace{1cm}&,&\hspace{1cm}\bar{\cal
B}_0=-\frac{d^2}{dx^2}+\frac{5}{2}
\end{eqnarray*}
govern the small fluctuations around the background
$\phi_{\bar{B}_\alpha}$. Thus,
\[
\Delta_1\varepsilon^{\rm K}(\alpha)=\frac{\hbar
m_d}{2}\lim_{s\rightarrow -\frac{1}{2}}
\left(\frac{\mu^2}{m_d}\right)^{s+\frac{1}{2}} [\zeta_{P\bar{\cal
K}}(s)-\zeta_{\bar{\cal B}_\alpha}(s)]
\]
is the Casimir kink energy renormalized with respect to the
$\phi_{\bar{B}_\alpha}$ background. From the asymptotic expansion
of both $\zeta_{P\bar{\cal K}}(s)$ and $\zeta_{\bar{\cal
B}_\alpha}(s)$ we obtain:
\[
\Delta_1\varepsilon^{\rm K}(\alpha)\cong \frac{\hbar
m_d}{2}\left\{\frac{2}{\Gamma(-\frac{1}{2})}+\lim_{s\rightarrow
-\frac{1}{2}}\left(\frac{2\mu^2}{5m_d^2}\right)^{s+\frac{1}{2}}\frac{c_1(\bar{\cal
K}_\alpha)}{\sqrt{4\pi}}\frac{\gamma[s+\frac{1}{2},\frac{5}{2}]}{\Gamma(s)}+\sum_{n=2}^{n_0-1}
\frac{c_n(\bar{\cal
K}_\alpha)}{\sqrt{4\pi}}\left(\frac{2}{5}\right)^{n-1}\frac{\gamma[n-1,\frac{5}{2}]}{\Gamma(-\frac{1}{2})}
\right\}
\]
where $c_n(\bar{\cal K}_\alpha)=a_n(\bar{\cal K})-a_n(\bar{\cal
B}_\alpha)$. The deviation from the exact result is:
\[
{\rm error}_1=\frac{\hbar
m_d}{2}\left[-\frac{1}{4\sqrt{\pi}}\left(b_{n_0,\bar{\cal
K}}(-\textstyle\frac{1}{2})-b_{n_0,\bar{\cal
B}_\alpha}(-\textstyle\frac{1}{2})\right)+B_{P\bar{\cal
K}}(-\textstyle\frac{1}{2})-B_{\bar{\cal
B}_\alpha}(-\textstyle\frac{1}{2})\right].
\]

In order to implement the mass renormalization prescription, we
assume that virtual mesons running on the loop of the tadpole
graph have a mass of $\frac{m}{\sqrt{2}}$ half of the time and a
mass of $\sqrt{2}m$ the other half-time on average. The
normal-order is thus prescribed for annihilation and creation
operators of mesons with $M=\frac{\sqrt{5}}{2}m$ mass; this
amounts to considering
\[
\delta m^2= \frac{1}{2m_d L}\zeta_{\bar{\cal
B}_0}(\textstyle\frac{1}{2})
\]
as the infinite quantity associated with the single divergent
graph of the system. Zeta function regularization plus the
asymptotic expansion tell us that the induced counter-term adds
\begin{eqnarray*}
\Delta_2\varepsilon^{\rm K}(\alpha)&=&\left<\psi_{\rm K}|H(\delta
m^2)|\psi_{\rm K} \right>-
\left<\psi_{B_\alpha}|H(\delta m^2)|\psi_{B_\alpha}\right>\\
&\cong&-\frac{\hbar m_d}{2\sqrt{4\pi}}c_1(\bar{\cal
K}_\alpha)\lim_{s\rightarrow-\frac{1}{2}}\left(\frac{2\mu^2}{5m_d^2}\right)^{s-\frac{1}{2}}
\frac{\gamma[s+\frac{1}{2},\frac{5}{2}]}{\Gamma(s)}
\end{eqnarray*}
to the one-loop correction to the link kink mass, whereas the
error is
\[
{\rm
error}_2=\lim_{L\rightarrow\infty}\frac{\hbar}{4L}c_1(\bar{\cal
K}_\alpha)B_{\bar{\cal B}_0}(\textstyle\frac{1}{2})\,.
\]
The sum of the contributions coming from the $s\rightarrow -\frac{1}{2}$
poles of $\Delta_1\varepsilon^{\rm K}(\alpha)$ and $\Delta_2\varepsilon^{\rm K}(\alpha)$ vanishes:
\[
\frac{\hbar m_d}{2}\frac{c_1(\bar{\cal
K}_\alpha)}{\sqrt{4\pi}}\lim_{s\rightarrow
-\frac{1}{2}}\left(\frac{2\mu^2}{5m_d^2}\right)^{s+\frac{1}{2}}\left(
\frac{\gamma[s+\frac{1}{2},\frac{5}{2}]}{\Gamma(s)}-
\frac{\gamma[s+\frac{1}{2},\frac{5}{2}]}{\Gamma(s)}\right)= 0\, .
\]
The choice of $M=\frac{\sqrt{5}}{2}m$ as a mass renormalization
parameter leads to exactly the same result that we encountered in
the more conventional systems with loop kinks and we end with the answer:
\[
\Delta M_{K_\alpha}\cong \hbar m[\Delta_0+D_{n_0}(\alpha)]
\]
where $\Delta_0=-\frac{1}{2\sqrt{2\pi}}$ and
\[
D_{n_0}(\alpha)=-\sum_{n=2}^{n_0-1} c_n(\bar{\cal K}_\alpha) d_n
=- \sum_{n=2}^{n_0-1} c_n(\bar{\cal K}_\alpha)\,
\left(\frac{2}{5}\right)^{n-1}\frac{\gamma[n-1,\frac{5}{2}]}{8
\sqrt{2} \pi\, }\,.
\]

The coefficients and the partial sums up to $n_0=11$ for
$\alpha=1$ are shown in the following Table {\small
\begin{center}
\begin{tabular}{|c|c|c|c|}
\\[-0.5cm] \hline
$n$ & $c_n (\bar{\cal K}_1)$ & $n_0-1$ & $D_{n_0}(1)$ \\ \hline 2
& -9.3750 & 2 & 0.0968454 \\ 3 & 10.9375 & 3 & 0.0617547 \\ 4 &
-10.2567 & 4 & 0.0786049  \\ 5 & 7.89397 & 5 & 0.0703349  \\ 6 &
-5.12392 & 6 & 0.0741904  \\ 7 & 2.86874 & 7 & 0.0725233  \\ 8 &
-1.40987 & 8 & 0.0731872 \\ 9 & 0.61636 & 9 & 0.0729439 \\ 10 &
-0.24186 & 10 & 0.0730259 \\ \hline
\end{tabular}
\end{center}
We find:
\[
\Delta M_{K_1}\cong \hbar m[\Delta_0+ D_{11}(1)]= -0.199471 \hbar
m+ 0.0730259 \hbar m = -0.126445
 \hbar m
\]
as the approximation to the kink Casimir energy measured with
respect to the Casimir energy of the $\phi_{\bar{B}_1}(x)$
background field configuration.

The choice of $\alpha=1$ is optimum in the sense that for smaller
values of $\alpha$ a tendency of the quantum correction towards
$-\infty$ is observed whereas for $\alpha$ greater than 1 the
tendency is toward $+\infty$. In Figure 4 , $\alpha=1$ is
identified as the inflexion point of a family that interpolates
between two background configurations with bad features: too
abrupt if $\alpha=\infty$ and too smooth if $\alpha=0$.

\begin{figure}[htbp]
\centerline{\psfig{file=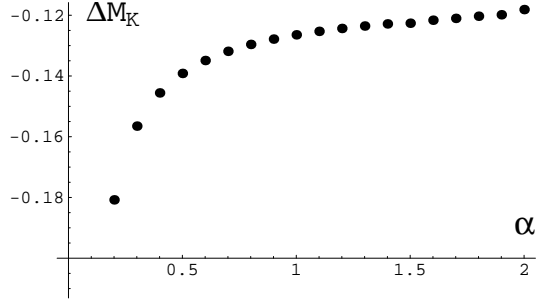, height=4.cm} } \caption{\small
\it Quantum correction to the kink mass as a function of $\alpha$
in the $(0.4,2.0)$ interval}
\end{figure}
We end this Section by comparing our renormalization criterion
with the prescription used in \cite{O'Brien}. Lohe and O'Brien
choose a mass renormalization parameter $M'$ in such a way that
the mass counter-term exactly cancels the difference in vacuum
Casimir energies between different points in the vacuum moduli
space.
\begin{equation}
\frac{\hbar m}{2\sqrt{2}}\left[\zeta_{\bar{\cal
V}_0}(-\textstyle\frac{1}{2})-\zeta_{\bar{\cal
V}_\pm}(-\textstyle\frac{1}{2})+\frac{3m}{\sqrt{2}}L\delta
m'^2\right]=0 \label{eq:ren}
\end{equation}
The contribution of the tadpole graph must be considered for
mesons with a suitable mass to satisfy (\ref{eq:ren}):
\[
\delta m'^2=\frac{1}{\sqrt{2}mL}\zeta_{\bar{\cal
V}}(\textstyle\frac{1}{2})\quad , \quad \bar{\cal
V}=-\frac{d^2}{dx^2}+\bar{M}'^2
\]
and we find $\bar{M}'^2=2.33$, $M'=\bar{M}'\frac{m}{\sqrt{2}}$, a
very close value to $M$. At the $L\rightarrow \infty$ limit
\[
\zeta_{\bar{\cal B}_0}(\textstyle\frac{1}{2})-\zeta_{\bar{\cal
V}}(\textstyle\frac{1}{2})=\frac{1}{4\pi}\log\frac{2,33}{2,50}
\]
If we had used $M'$ as the mass renormalization parameter, the
result would differ by
\[
\Delta_{K_1}M'-\Delta_{K_1}M=\frac{\hbar
m}{4\sqrt{2}}\frac{c_1(\bar{\cal
K}_1)}{(4\pi)^{\frac{3}{2}}}\log\frac{2.33}{2.50}
\]
which is a very small quantity indeed.

\section{Outlook}
The natural continuation of this work, and the main motivation to
develop the asymptotic method, is the computation of quantum kink
masses in theories with N-component scalar fields. Nevertheless,
explorations in the supersymmetric world along these lines are
also interesting.

All the models that we have described admit a supersymmetric
extension because the potential energy density always can be
written as $U(\psi)=\frac{1}{2}\frac{dW}{d\psi}\frac{dW}{d\psi}$.
In non-dimensional variables the superpotential $\bar{W}(\phi)$
for each model is:
\[
\bar{W}(\phi)=\pm 4{\rm
cos}\frac{\phi}{2}\hspace{0.5cm},\hspace{0.5cm}
\bar{W}(\phi)=\pm(\phi-\frac{\phi^3}{3})
\]
\[
\bar{W}(\phi)=\pm 4\frac{1}{\sqrt 2}(\frac{1}{4}{\rm sinh}2\phi
-\frac{3}{2}\phi ) \hspace{0.5cm},\hspace{0.5cm}
\bar{W}(\phi)=\pm\frac{\phi^2}{2}(\frac{\phi^2}{2}-1)\,.
\]
The supersymmetric extension includes also a Majorana spinor
field:
\[
\chi(x^\mu)=\left(\begin{array}{c}\chi_1(x^\mu)\\\chi_2(x^\mu)\end{array}\right)\hspace{0.5cm},
\hspace{0.5cm}\chi_\alpha^*=\chi_\alpha \quad ,\quad \alpha=1,2\,.
\]
Choosing the Majorana representation $\gamma^0=\sigma^2,
\gamma^1=i\sigma^1, \gamma^5=\sigma^3$ of the Clifford algebra
$\{\gamma^\mu,\gamma^\nu\}=2g^{\mu\nu}$ and defining the Majorana
adjoint $\bar{\chi}=\chi^t \gamma^0$, the action of the
supersymmetric model is:
\[
S=\frac{1}{2c_d^2}\int
dx^2\left\{\partial_\mu\phi\partial^\mu\phi+i\bar{\chi}\gamma^\mu\partial_\mu\chi
-\frac{d\bar{W}}{d\phi}\frac{d\bar{W}}{d\phi}-\bar{\chi}\frac{d^2\bar{W}}{d\phi^2}\chi\right\}\,.
\]
The ${\cal N}=1$ supersymmetry transformation is generated on the
space of classical configurations by the Hamiltonian spinor
function
\[
Q=\int
dx\left\{\gamma^\mu\gamma^0\chi\partial_\mu\phi+i\gamma^0\chi\frac{d\bar{
W}}{d\phi} \right\}\,.
\]
The components of the Majorana spinorial charge Q close the
supersymmetry algebra
\begin{equation}
\{Q_\alpha,Q_\beta\}=2(\gamma^\mu\gamma^0)_{\alpha\beta}P_\mu-2\gamma^1_{\alpha\beta}T
 . \label{alg}
\end{equation}
Their (anti)-Poisson bracket is given in (\ref{alg}) in terms of
the momentum $P_\mu$ and the topological central charge $T=|\int
d\bar{W}|$ .

The chiral projections $Q_{\pm}=\frac{1\pm\gamma^5}{2}Q$ and
$\chi_{\pm}=\frac{1\pm\gamma^5}{2}\chi$ provide a very special
combination of the supersymmetric charges:
\[
Q_++Q_-=\int
dx\left\{(\chi_+-\chi_-)\frac{d\phi}{dx}-(\chi_-+\chi_+)\frac{d\bar{W}}{d\phi}\right\}\,.
\]
$Q_++Q_-$ is zero for the classical configurations that satisfy
$\frac{d\phi}{dx}=\mp\frac{d\bar{W}}{d\phi}$ and $\chi_{\pm}=0$
which are thus classical BPS states. One immediately notices that
our kinks are such BPS states and besides the small bosonic
fluctuations one must take into account the small fermionic
fluctuations around the kink for computing the quantum correction
to the kink mass in the extended system. The fermionic
fluctuations around the kink configuration lead to other solutions
of the field equations if the Dirac equation
\[
\left\{i\gamma^\mu
\partial_\mu+\frac{d^2\bar{W}}{d\phi^2}(\phi_{\rm K}
) \right\}\delta_F\chi(x,t)=0
\]
is satisfied. We multiply this equation for the adjoint of the
Dirac operator
\[
\left\{-i\gamma^\mu\partial_\mu+\frac{d^2\bar{W}}{d\phi^2}(\phi_{\rm
K})\right\}
\left\{i\gamma^\mu\partial_\mu+\frac{d^2\bar{W}}{d\phi^2}(\phi_{\rm
K})\right\}\delta_F\chi(x,t)=0
\]
and, due to the time-independence of the kink background, look for
solutions of the form: $\delta_F(x,t)=f_F(x;\omega)e^{i\omega t}$.
This is tantamount to solving the spectral problem
\[
\left\{-\frac{d^2}{dx^2}+\frac{d^2\bar{W}}{d\phi^2}(\phi_{\rm
K})\frac{d^2\bar{W}}{d\phi^2}(\phi_{\rm K}) \mp
i\gamma^1\frac{d\bar{W}}{d\phi}(\phi_{\rm
K})\frac{d^3\bar{W}}{d\phi^3}(\phi_{\rm K})\right\}
f_F(x;\omega)=\omega^2f_F(x;\omega)\,.
\]
Projecting onto the eigen-spinors of $i\gamma^1$,
\[
f_F^{(1)}(x;\omega)=\frac{1+i\gamma^1}{2}f_F(x;\omega)=\frac{1}{2}
\left(\begin{array}{c}f_F^+(x;\omega)-f_F^-(x;\omega)\\-f_F^+(x;\omega)+f_F^-(x;\omega)
\end{array}\right)
\]
we end with the spectral problem:
\[
\left\{-\frac{d^2}{dx^2}+\frac{d^2\bar{W}}{d\phi^2}(\phi_{\rm
K})\frac{d^2\bar{W}}{d\phi^2}(\phi_{\rm K}) \mp
\frac{d\bar{W}}{d\phi}(\phi_{\rm
K})\frac{d^3\bar{W}}{d\phi^3}(\phi_{\rm K})\right\}
f_F^{(1)}(x;\omega)={\cal
K}f_F^{(1)}(x;\omega)=\omega^2f_F^{(1)}(x;\omega)
\]
for the same Schr\"odinger operator as that governing the bosonic
fluctuations.

Therefore, generalized zeta function methods can also be used in
supersymmetric models for computing the quantum corrections to the
mass of BPS kinks. Great care however, is needed in choosing the
boundary conditions on the fermionic fluctuations without spoiling
supersymmetry. We look forward to extend this research in this
direction.\\\\\\
{\Large\bf Acnowledgments}\\\\
W. G. F. acknowledges the partial financial support to his work provided by DGICYT under contract BFM 2000 0357.\\

\section*{Appendix}
In this Appendix we describe the iterative procedure that gives
the coefficients $a_n(x,x)$ used in the text. For alternative
descriptions, see \cite{DeWitt}, \cite{Stone}. For an interesting
interpretation of these coefficients as invariants of the
Korteweg-de Vries equation, see \cite{Perelomov}.

Starting from formula (\ref{eq:pde}) in the text, we write the recurrence relation
\begin{equation}
(n+1) \, a_{n+1}(x,y)+(x-y) \frac{\partial a_{n+1}(x,y)}{\partial
x}- \bar{V}(x) a_n(x,y)=\frac{\partial^2 a_n(x,y)}{\partial
x^2}. \label{eq:recu8}
\end{equation}
In order to take the limit $y\rightarrow x$ properly, we introduce the notation
\[
{^{(k)}A}_n(x)=\lim_{y \rightarrow x} \frac{\partial^k
a_n(x,y)}{\partial x^k}
\]
and, after differentiating (\ref{eq:recu8}) $k$ times, we find
\[
{^{(k)} A}_n(x) =\frac{1}{n+k} \left[ \rule{0cm}{0.6cm} \right.
{^{(k+2)} A}_{n-1}(x) + \sum_{j=0}^k {k \choose j}
\frac{\partial^j \bar{V}(x)}{\partial x^j}\, \, {^{(k-j)}
A}_{n-1}(x) \left. \rule{0cm}{0.6cm} \right].
\]
from this equation and ${^{(k)} A}_0(x)=\lim_{y\rightarrow x}
\frac{\partial^k a_0}{\partial x^k}= \delta^{k0}$, all the
${^{(k)} A}_n(x)$ can be generated recursively. Returning to
(\ref{eq:recu8}), we finally obtain a well-defined recurrence
relation
\[
a_{n+1}(x,x)=\frac{1}{n+1} \left[ {^{(2)} A}_n(x)+\bar{V}(x)
\, a_n(x,x) \right]
\]
suitable for our purposes.

We give the explicit expressions of the first eight $a_n(x,x)$
coefficients. The abbreviated notation is
$u_k=\frac{d^k\bar{V}}{dx^k}(x)$,
$u_k^n=\left(\frac{d^k\bar{V}}{dx^k}(x)\right)^n$ :
\begin {eqnarray*}
a_1(x,x)&=&u_0 \\\\ a_2(x,x)&=&\frac{1}{2}u_0^2 +
\frac{1}{6}u_2\\\\ a_3(x,x)&=&\frac{1}{6} u_0^3 + \frac{1}{6}u_2
u_0 + \frac{1}{12} u_1^2 +
  \frac{1}{60} u_4\\\\
a_4(x,x)&=&\frac{1}{24}u_0^4 + \frac{1}{12}u_2 u_0^2 +
  \frac{1}{12} u_1^2 u_0 + \frac{1}{60}u_4 u_0 +
  \frac{1}{40}u_2^2+ \frac{1}{30} u_1u_3 +
  \frac{1}{840} u_6\\\\
a_5(x,x)&=&\frac{1}{120} u_0^5 + \frac{1}{36} u_2 u_0^3 +
  \frac{1}{24}u_1^2 u_0^2 + \frac{1}{120} u_4 u_0^2 +
  \frac{1}{40}u_2^2 u_0 +
  \frac{1}{30} u_1u_3 u_0+
  \frac{1}{840} u_6u_0+
  \frac{11}{360}u_1^2 u_2\\&+&
  \frac{23}{5040}u_3^2 +
  \frac{19}{2520}u_2 u_4+
  \frac{1}{280} u_1u_5 +\frac{1}{15120} u_8\\\\
a_6(x,x)&=&\frac{1}{720}u_0^6 + \frac{1}{144} u_2u_0^4 +
  \frac{1}{72} u_1^2 u_0^3 + \frac{1}{360} u_4u_0^3+
  \frac{1}{80} u_2^2 u_0^2 +
  \frac{1}{60} u_1u_3 u_0^2+
  \frac{11}{360} u_1^2 u_2 u_0+
  \frac{1}{280} u_1u_5 u_0\\&+&
  \frac{1}{288} u_1^4 + \frac{1}{15120}u_8 u_0 +
  \frac{61}{15120}u_2^3+
  \frac{43} {2520}u_1u_2 u_3 +
  \frac{23}{5040} u_0 u_3^2+
  \frac{5}{1008}u_1^2 u_4+
  \frac{19}{2520}u_0 u_2 u_4\\ &+&
  \frac{23}{30240} u_4^2+
  \frac{19}{15120} u_3 u_5+
  \frac{1}{1680}u_0^2u_6+
  \frac{11}{15120} u_2 u_6+
  \frac{1}{3780}u_1u_7 + \frac{1}{332640}u_{10}\\\\
a_7(x,x)&=&\frac{1}{5040}u_0^7 + \frac{1}{720} u_2u_0^5 +
  \frac{1}{288} u_1^2 u_0^4 + \frac{1}{240} u_2^2 u_0^3+
  \frac{1}{180} u_1 u_3 u_0^3+
  \frac{11}{720} u_1^2 u_2 u_0^2 +
  \frac{1}{560} u_1 u_5 u_0^2\\&+&
  \frac{1}{288} u_1^4 u_0+ \frac{61}{15120} u_2^3 u_0+
  \frac{43}{2520} u_1 u_2 u_3u_0 + \frac{5}{1008} u_1^2 u_4u_0 +
  \frac{1}{332640}u_{10} u_0 +
  \frac{23}{10080} u_3^2 u_0^2\\&+&
  \frac{19}{5040} u_2 u_4 u_0^2+
  \frac{1}{5040}u_6 u_0^3+
  \frac{83}{10080}u_1^2 u_2^2 +
  \frac{1}{252} u_1^3u_3+
  \frac{31}{10080} u_2 u_3^2+
  \frac{1}{280} u_1 u_3 u_4+
  \frac{1}{1440}u_0^4 u_4\\ &+&
  \frac{5}{2016} u_2^2 u_4+
  \frac{23}{30240} u_0 u_4^2+
  \frac{1}{420} u_1 u_2 u_5+
  \frac{19}{15120} u_0 u_3u_5 +
  \frac{71}{665280} u_5^2+
  \frac{1}{2016}u_1^2 u_6\\&+&
  \frac{11}{15120} u_0 u_2 u_6+
  \frac{61}{332640}u_4u_6+
  \frac{1}{3780}u_0 u_1 u_7+
  \frac{19}{166320} u_3u_7 +
  \frac{1}{30240}u_0^2 u_8+
  \frac{17}{332640}u_2u_8\\ &+&
  \frac{1}{66528}u_1 u_9+
  \frac{1}{8648640}u_{12}\\\\
a_8(x,x)&=&\frac{1}{40320}u_0^8 + \frac{1}{960} u_2^2 u_0^4 +
  \frac{1}{720} u_1 u_3 u_0^4 +
  \frac{1}{576} u_1^4 u_0^2 +
  \frac{1}{252} u_1^3 u_3 u_0+
  \frac{1}{280} u_1 u_3 u_4 u_0+
  \frac{1}{420} u_1 u_2 u_5 u_0\\ &+&
  \frac{31}{10080} u_2u_3^2 u_0 +
  \frac{5}{2016} u_2^2 u_4 u_0+
  \frac{1}{2016}u_1^2 u_6 u_0+
  \frac{1}{8648640}u_{12} u_0 +
  \frac{23}{60480} u_4^2 u_0^2 +
  \frac{19}{30240} u_3 u_5 u_0^2\\ &+&
  \frac{11}{30240}u_2u_6 u_0^2+
  \frac{1}{7560}u_1 u_7u_0^2+
  \frac{11}{2160} u_1^2 u_2 u_0^3 +
  \frac{1}{90720}u_8 u_0^3+ \frac{1}{7200}u_4 u_0^5+
 \frac{1}{1440}u_0^5 u_1^2\\ &+& \frac{1}{4320}u_0^6 u_2+
  \frac{17}{8640} u_1^4 u_2+
  \frac{83}{10080} u_0 u_1^2 u_2^2+
  \frac{61}{30240} u_0^2 u_2^3+
  \frac{1261}{1814400} u_2^4+
  \frac{43}{5040} u_0^2 u_1u_2u_3\\&+& \frac{227}{37800} u_1 u_2^2
    u_3+ \frac{23}{30240} u_0^3 u_3^2 +
  \frac{659}{302400} u_1^2 u_3^2+
  \frac{5}{2016} u_0^2 u_1^2 u_4+
  \frac{19}{15120} u_0^3 u_2u_4+
  \frac{527}{151200} u_1^2 u_2u_4\\&+& \frac{7939}{9979200} u_3^2u_4+
  \frac{6353}{9979200}u_2u_4^2+
  \frac{1}{1680}u_0^3 u_1 u_5+
  \frac{17}{30240} u_1^3 u_5+
  \frac{13}{12320}u_2u_3u_5 +
  \frac{3067}{4989300} u_1 u_4u_5\\&+& \frac{71}{665280} u_0 u_5^2+
  \frac{1}{20160}u_0^4 u_6+
  \frac{3001}{9979200} u_2^2u_6 +
  \frac{13}{29700} u_1 u_3u_6+
  \frac{61}{332640} u_0 u_4u_6\\&+&
  \frac{3433}{259459200} u_6^2+
  \frac{109}{498960} u_1 u_2u_7 +
  \frac{19}{166320} u_0 u_3u_7 +
  \frac{1501}{64864800} u_5u_7+
  \frac{71}{1995840} u_1^2 u_8\\&+&
  \frac{17}{332640} u_0 u_2 u_8 +
  \frac{2003}{129729600} u_4u_8 +
  \frac{1}{66528}u_0 u_1 u_9+
  \frac{5}{648648} u_3u_9+
  \frac{1}{665280}u_0^2u_{10}\\&+&
  \frac{73}{25945920} u_2u_{10}+
  \frac{1}{1441440}u_1 u_{11}+
  \frac{1}{259459200}u_{14}
\end {eqnarray*}

\end{document}